# Method to extracting the penetration field in superconductors from DC magnetization data

Evgueni F. Talantsev[1, *,**]

[1]M.N. Mikheev Institute of Metal Physics, Ural Branch, Russian Academy of Sciences, 18, S. Kovalevskoy St., Ekaterinburg, 620108, Russia

*corresponding author

**corresponding author's E-mail: evgeny.talantsev@imp.uran.ru

## Abstract

The lower critical field, $B_{c1}$, is one of the fundamental quantities of a superconductor which directly manifests the Cooper pair bulk density in the material. Although this field can be measured using several techniques, the most conventional method is to calculate this field from the experimentally measured DC penetration field, $B_p$, which is defined as the starting point of the deviation of the DC magnetization curve, $M(B_{appl})$, from a linear dependence. Surprisingly, we found no mathematical routine which describes how this starting point of deviation can be found. Here, we propose the extraction of $B_p$ from the fit of $M(B_{appl})$ dataset to the power law, where the threshold criterion $M_c$ can be established by a convention. The advantage of this approach is that the procedure extracts one additional characteristic parameter: the power-law exponent. We demonstrated the applicability of this approach to polycrystalline $ThIr_3$, $WB_{4.2}$, $BaTi_2Bi_2O$, $Th_4H_{15}$, to thin films of Pb and $MgB_2$ and to Nb single crystal. In most reports, $B_{c1}(T)$ analysis is limited by the extraction of the London penetration depth. We advanced the analysis to extract primary thermodynamic superconducting parameters (i.e. the ground state superconducting energy gap, $\Delta(0)$, the relative jump in electronic specific heat at transition temperature, $\frac{\Delta C}{\gamma T_c}$, and the gap-to-transition temperature ratio, $\frac{2\Delta(0)}{k_B T_c}$) from $B_{c1}(T)$ data. This extraction was performed for Nb, $ThIr_3$, $TaRh_2B_2$ and $NbRh_2B_2$.

**Method to extracting the penetration field in superconductors from DC magnetization data**

## I. Introduction

Walther Meissner and Robert Ochsenfeld discovered that an external magnetic field, $B$, is expelled from the superconducting tin and lead [1]. The maximum magnetic flux density at which the superconducting state starts to collapse [2] described by following equations:

$$B_c(T) = \frac{\phi_0}{2 \cdot \sqrt{2} \cdot \pi} \cdot \frac{1}{\lambda(T) \cdot \xi(T)}, \text{ for Type-I superconductors} \quad (1)$$

$$B_{c1}(T) = \frac{\phi_0}{4 \cdot \pi} \cdot \frac{ln\left(\frac{\lambda(T)}{\xi(T)}\right) + 0.5}{\lambda^2(T)}, \text{ for Type-II superconductors} \quad (2)$$

where λ is the London penetration depth, ξ is the coherence length, and $\phi_0 = \frac{h}{2 \cdot e} \approx 2.07 \cdot 10^{-15}\ Wb$ is the superconducting flux quantum, where $h$ is the Planck constant, and $e$ is the electron charge. The effect of magnetic flux expelling from superconductors is recognized as one of the most fundamental effects in superconductivity [3-10], which has been utilized in several superconducting technologies [11-16].

It should be stressed that both Eqs. 1,2 can be represented by a universal equation [17,18]:

$$B_{MO}(T) = \frac{\phi_0}{4\pi} \frac{ln\left(1 + \sqrt{2} \cdot \frac{\lambda(T)}{\xi(T)}\right)}{\lambda^2(T)} = \frac{\phi_0}{2\pi} \frac{\mu_0 \cdot e^2}{m_e} n(T) \times ln\left(1 + \sqrt{2} \frac{\lambda(T)}{\xi(T)}\right), \quad (3)$$

where the subscript MO designates the Meissner-Ochsenfeld field (i.e. the maximum flux density at which the superconducting state starts to collapse), $m_e$ is the mass of the charge carrier, $\mu_0$ is magnetic permeability of free space, and the bulk Cooper pair density describes by the following equation:

$$n(T) = \frac{1}{2} \frac{m_e}{\mu_0 e^2 \lambda^2(T)} \quad (4)$$

(a detailed discussion of the advances of Eq. 3 vs Eqs. 1,2 presented recently [19]). Based on Eqs. 3,4, the measurement of $B_{MO}(T)$, that is the maximum expelled magnetic flux density in the superconductor, is a direct way to determine the bulk density of Cooper pairs, $n(T)$, in the superconductor.

One of the most conventional methods for determining $B_{MO}$ is based on the analysis of $M(B_{appl})$ curve to find a point where the curve deviates from a straight line (this line is also designated as the Meissner line). Perkins *et al.* [20] and Eley *et al.* [21] designated this starting point of the deviation as the penetration field, $B_p$. This designation is commonly accepted in the field; however, some research groups [22] used different notation for this field. Based on the sample geometry, the determined $B_p$ is used to calculate $B_{MO}$. To calculate this field for a disk, exact expression proposed by Brandt can be used [23]:

$$B_{MO} = \frac{B_p^{disk}}{1-N} = \frac{B_p^{disk}}{1-\left(1-\frac{1}{1+q\times\frac{a}{b}}\right)} \quad (5)$$

$$q = \frac{4}{3\pi} + \frac{2}{3\pi} tanh\left(1.27\frac{b}{a} ln\left(1+\frac{a}{b}\right)\right) \quad (6)$$

where $2a$ is the strip width or the disk diameter and $2b$ is the sample thickness. It should be mentioned that there is an alternative approach, where the calculation is based on the demagnetization factor, $N$ (which is another way to account the sample geometry, more details can be founded in Ref. 23).

Based on that, accurate determination of $B_{MO}$ is directly related to accurate extraction of the first flux entry field, $B_{en}$, from experimental $M(B_{appl})$ curves.

Surprisingly, we found that there is no standard procedure for this extraction. Available procedure descriptions are pretty unclear: "… *At low magnetic fields (H < 15 Oe) the* $M_v$ *data collected at 0.5 K was fitted with the linear formula* $M_v(H) = -bH$*, where b is the slope of the fitted line. … A rough estimation of the lower critical field at 0.5 K (not corrected for demagnetization) is* $H_{c1}(0.5\ K) = 10$ *Oe.*" [24] (in this paper Carnicom *et al* studied $WB_{4.2}$ phase).

This description should be commented upon, because if the fit of the volume magnetization, $M_v$, to linear function was performed in applied magnetic fields ($B_{appl}$) of 0 mT < $B_{appl}$ < 1.5 mT, how it was possible to define the first flux entry field as $B_p$(0.5 K) = 1.0 mT, because, by the definition, $B_p$ is the starting point for the deviation of $M_v(B_{appl})$ from straight line. And, why is the field range of 0 mT < $B_{appl}$ < 1.5 mT was chosen for the linear fit? Similar uncertainties of the used procedure and, what is important, unexplained arbitrary chosen field range, which is used for the linear fit, can be found in hundreds of papers.

In this paper we proposed strict mathematical procedure to extract the first flux entry field from the magnetization data and demonstrated the applicability of the proposed technique for a wide range of superconductors: for pure metals (type-I lead and type-II niobium), $Th_4H_{15}$ hydride, $BaTi_2Bi_2O$ oxide, boron-based bulk $WB_{4.2}$, , $TaRh_2B_2$ and $NbRh_2B_2$ compounds and $MgB_2$ thin film, and 5*d*-electron compound$ThIr_3$.

## II. Model description

In this work we propose the procedure for extracting the penetration field from the magnetization data, $M(B_{appl})$, which is similar to that used to deduce the transport critical current, $I_c$, from the $E(I)$ curves [25]:

$$E(I) = E_0 + k \times I + E_c \times \left(\frac{I}{I_c}\right)^n \qquad (7)$$

where $E_0$ is the instrumental offset, $k$ is a linear term used to accommodate incomplete current transfer in short samples, $E_c$ is the electric field criterion (for which the conventional value is $E_c$ = 1 μV/cm [25,26]) and $n$ is the power-law exponent. It should be noted that $E_0$, $k$, and $n$ are free-fitting parameters, while $E_c$ is arbitrary chosen fixed parameter.

Despite the fact that the critical current, $I_c$, can be defined based on fundamentally different physical effect [27,28], the fit of $E(I)$ curves to the power law (Eq. 7) remains the most conventionally used routine to deduce the critical currents from transport current measurements [26].

The power-law approximation of $E(I)$ curves (Eq. 7) is reasonably accurate [26-28] if the analysed $E(I)$ dataset has its amplitude, $E(I)_{max}$, not much exceeded the $E_c$ criterion:

$$E(I) - E_0 - k \times I \lesssim p \times E_c \quad (8)$$

where $E_0$ and $k$ are the free-fitting parameters in Eq. 7, and $p \lesssim 5$. It should be noted, that because the primary purpose of using Eqs. 7,8 is to determine the parameters at which the dissipation starts, that lower $p$ value implies that more accurately the parameters in Eqs. 7,8 will be deduced. For instance, we can mention recent report by Yanagisawa *et al* [29] where measurements and data fits were performed for $p \lesssim 1$.

By applying several different fitting functions to approximate $M(B_{appl})$ curves, measured as for Type-I, as for Type-II superconductors, we found that the power-law function is the one which is most simple and accurate. Thus, to fit $M(B_{appl})$ curves, we propose to use the fitting function:

$$M(B_{appl}) = M_0 + k \times B_{appl} + M_c \times \left(\frac{B_{appl}}{B_p}\right)^n \quad (9)$$

where $M_0$ is an instrumental offset, $k$ is a linear term (which is also named Meissner slope), and $M_c$ is the threshold criterion (which we discuss below) and $n$ is the power-law exponent.

It should be emphasized that there is a close approach that defines $B_p$ based on the following equation:

$$M(B_p) - k \times B_p = M_c \quad (10)$$

For instance, Eq. 10 was used by Eley *et al* [21] to deduce $B_p(T)$ and $B_{c1}(T)$ in $HgBa_2CuO_{4+\delta}$ single crystal. For the definition of $B_p(T)$, Eley *et al* [21] utilized the criterion:

$$M_c = 0.005 \times \left|M(B_{appl}, T_{lowest})\right|_{max} \quad (11)$$

where $M(B_{appl}, T_{lowest})$ is full DC magnetization curve measured at the lowest experimentally available temperature, $T_{lowest}$ (which was $T_{lowest} = 10\ K$ in [23]). However, in the literature, a considerably less strict $M_c$ criterion:

$$M_c = 0.08 \times |M(B_{appl}, T_{lowest})|_{max} \tag{12}$$

is in a use also [30].

The best practice for $M(B_{appl})$ data fit to Eq. 9 is to utilize strict $M_c$ criterion defined by Eq. 11. However, this strict $M_c$ criterion may not be possible to use, if $M(B_{appl})$ data was measured for samples with $B_p \lesssim 2\ mT$ by conventional vibrating sample magnetometers (VSM), which typically have $B_{appl}$ step limit of 0.2 mT (see, for instance Fig. 3 in Ref. 24). Because Eq. 9 has four free-fitting parameters (i.e., $M_0$, $k$, $B_p$ and $n$) the $M(B_{appl})$ dataset should have at least 10 datapoints to extract parameters with reasonable accuracy. Primary condition for the choice of appropriated $M_c$ criterion for given sample and the measuring apparatus is based on the number of $M(B_{appl})$ data points which can be measured before the datapoints start to departure from the Meissner line. If there are not many datapoints in the Meissner line, then much less strict criterion (for instance, $M_c = 0.02 \times |M(B_{appl}, T_{lowest})|_{max}$) should be used. This choice of $M_c$ makes it possible to fit larger set of raw $M(B_{appl})$ data points to Eq. 9, because $p$-value (in Eqs. 8,13) will be increased too.

This implies that less strict criterion (for instance, $M_c = 0.02 \times |M(B_{appl}, T_{lowest})|_{max}$) should be used to analyse $M(B_{appl})$ datasets for superconductors with $B_p \lesssim 2\ mT$ (if the material is studied in standard VSM apparatus with $B_{appl}$ step of 0.2 mT). Alternatively, measurements can be performed in the superconducting quantum interreference vibrating sample magnetometers (SQUID VSM). These machines typically have $B_{appl}$ step limit of ~ 0.03 mT. The use of SQUID VSM is also preferable option for studies of "metallurgically realistic" superconductors [31] which have smooth $M(B_{appl})$ curves and, thus, the determination of the initial deviation point from the Meissner line is a challenging problem. It should be also noted, that utilized, in some works, a criterion based on visually chosen deviation point from the Meissner line very often leads to the overestimation of the deduced $B_p$ .

As mentioned above, the power-law function (Eq. 7) is a good approximation of the $E(I)$ curves if Eq. 8 is satisfied. Another condition is that $M(B_{appl})$ curves should be measured at narrow $B_{appl}$ steps to make it possible to deduce $B_p$ with high accuracy. This particularly true for samples in which the power-law exponent $n$ (Eq. 9) was not high, $n \lesssim 5$.

Considering the different options under similar conditions for the $M(B_{appl})$ data fitted to Eq. 9, we found that the conditions described in Eq. 7 can be also applied to the $M(B_{appl})$ data fit:

$$M(B_{appl}) - M_0 - k \times B_{appl} \lesssim p \times M_c \quad (13)$$

where preferable value for $p \lesssim 5$.

## III. Results and Discussion

### 3.1. Polycrystalline $ThIr_3$

To demonstrate the main features of the proposed routine (Eqs. 9,11,13) in Figures 1,2 we showed the $M(B_{appl}, T = 1.7\ K)$ data for polycrystalline $ThIr_3$ sample (data reported by Górnicka *et al* [32]) together with data fit to Eqs. 9. In Figure 1, the full $M(B_{appl}, T = 1.7\ K)$ curve is shown, where $|M(B_{appl}, T = 1.7\ K)|_{max}$ and $B_{M,max}$ are shown. In Figure 2, the evolution of the deduced parameters vs. the range of the used $B_{appl}$ is shown.

As shown in Figure 2, there is a minimal $M(B_{appl}, T)$ dataset for which reliable data fitting can be performed. For instance, in Figure 2, we highlight a group of deduced values obtained by the data fit for the low $B_{appl}/B_{M,max}$ range.

### 3.2. Polycrystalline $WB_{4.2}$

We now return to the above-mentioned above paper by Carnicom *et al* [24], who reported $M_V(B_{appl}, T)$ for $WB_{4.2}$ ceramic. In Figures 3,4 we showed $M_V(B_{appl}, T = 0.5\ K)$ data and fit to Eqs. 9,11,13. From this, it is clear that accurate determination of $B_p$ requires raw $M_V(B_{appl}, T)$ measured with small steps, especially at $B_{appl}$. For the given $M_V(B_{appl}, T = 0.5\ K)$ dataset, the deduced $B_p$ has a relative uncertainty of 20%, if:

$$\frac{M(B_{appl}) - M_0 - k \times B_{appl}}{M_c} \leq 8 \tag{14}$$

The uncertainties of the deduced parameters can be reduced by choosing wider $B_{appl}$ range for the fit (see Figure 4):

$$\frac{M(B_{appl}) - M_0 - k \times B_{appl}}{M_c} \leq 35. \tag{15}$$

However, there can be an issue with the accuracy of the deduced parameters, if a larger *p*-value (Eq. 13) was used.

### 3.3. Polycrystalline $BaTi_2Bi_2O$

To demonstrate that $B_p$ and *n*-value and their uncertainties can be reliably deduced for the $M(B_{appl}, T)$ datasets which have dense raw data at low $B_{appl}$ in Figures 5 and 6 we analysed $M(B_{appl}, T = 1.85\ K)$ reported by Yajima [33] for polycrystalline $BaTi_2Bi_2O$. The fitting parameters $B_p, n, k$ (Figures 5 and 6) vary within narrow intervals, and an increase in $B_{appl}$ range (used for the analysis) causes a reduction in the uncertainties.

### 3.4. Polycrystalline $Th_4H_{15}$

The fourth representative ceramic is shown in Figure 7, where the analysis was performed on the magnetization curve of $Th_4H_{15}$ measured at $T$ = 2 K (raw data reported by Wang *et al* [34]).

Overall, $M\left(B_{appl},T\right)$ data analysis for ceramic samples (Figures 1-7) shows that these datasets have wide transitions, which can be characterized by $n \leq 4$ and, thus, it is a challenging to deduce $B_p$ (i.e. the point where the $M\left(B_{appl}\right)$ curve starts to deviate from the straight line) for such smooth curves using an eye-guided approach.

In the subservient sessions we applied Eqs. 9,11,13 to analyse the data for thin films of lead (Section 3.5) and bulk single crystals of niobium (Section 3.6), where the $M\left(B_{appl},T\right)$ curves demonstrate sharp transitions, which are characterized by $n > 10$.

### 3.5. Lead thin films deposited on mica substrate

Lock [35] reported $M\left(B_{appl},T = 4.2\ K\right)$ data for four lead thin films deposited on mica substrate. Raw data is shown in Figure 8 together with the fits to Eqs. 9,11,13. The results for $B_p$ and $n$ are summarised in Fig. 9.

Figs. 8,9 represent a nice demonstration for films granularity problem which can be, in more general, considered as a demonstration of the problem defined Matthias [31] as "metallurgically realistic Type III superconductors" problem in terms of analysis of the $M\left(B_{appl},T\right)$ data. Despite a fact that term of "Type-III superconductors" did not become widely used, it describes samples of both Type-I and Type-II superconductors, in which various types of defects and imperfections smoothing the $M\left(B_{appl},T\right)$ curve.

The meaning of $n$-value can be understood, if one considers three lead films with thickness $2b$ > 180 nm, which all of which exhibit $B_p \sim 60\ mT$ (which all of which are in the expected range [2,36]) and $n > 10$. However, further thinning of the film to $2b$ = 107 nm (Fig. 8(d) and Fig. 9) caused dramatic drops at $B_p = 38\ mT$ and $n = 3.2$. The explanation of this result is that 107 nm thick film has a granular structure, where grain boundaries represent the weakest part of the film, and the applied magnetic field penetrates at a much lower $B_{en}$ in comparison with the field required to penetrated inside of the grains or solid sample, such as films with $2b > 180$ nm.

Based on this, the concept of "Type-III superconductors" [31] can be replaced by a more precisely defined (based on mathematical analysis of experimental data) concept of *n*-value for the fit of $M(B_{appl}, T)$ curve to Eq. 9.

### 3.6. $MgB_2$ epitaxial thin film deposited on SiC substrate

To demonstrate that proposed method is also applicable for epitaxial thin films of type-II superconductors, in Fig. 10 we show $M(B_{appl}, T = 5\ K)$ data and data fit to Eq. 9 for 300 nm thick epitaxial $MgB_2$ deposited on SiC substrate (raw data reported by Tan *et al.* [37]). For the fit, we utilized the criterion described by Eq.11, which implies that $M_c$ = 0.004 A/m. The results for $B_p$ and $n$ are shown in Fig. 10.

### 3.7. $M(B_{appl}, T)$ and $B_{c1}(T)$ data analysis for niobium single crystal

Niobium is a practically important low-$\kappa$ Type-II superconductor which exhibits $\kappa \cong 1$ [2,38,39]. This $\kappa$ is near the lower limit of $\kappa = \frac{1}{\sqrt{2}}$ associated with Type-II superconductors.

Stromberg [39] reported detailed magnetization studies for high-purity niobium single crystals, from which in Figure 10 we show several $M(B_{appl}, T)$ datasets measured at different temperatures for Sample Nb-49 together with the fits to Eqs. 9,11,13. For this sample, we use $M_c$ criterion of:

$$M_c = 0.02 \times M(B_{appl} = 163.0\ mT, T = 1.154\ K) = 0.02 \times 9.95\ \frac{kA}{m} = 199\ \frac{A}{m}. \quad (16)$$

The deduced temperature-dependent $B_p(T)$ and $n(T)$ are shown in Fig. 12. All deduced $n > 18$, which implies that the studied sample of Nb-49 exhibits a nearly perfect structure. In Fig. 12 deduced *n*-values are shown in linear-log plot. This type of plot is commonly used in applied superconductivity when experimental data is demonstrated at wide temperature range of $0 < T \leq T_c$ [21,40,41].

**Figure 12.** Temperature dependent $B_p(T)$ and $n(T)$ deduced for single crystal of niobium (sample Nb-49) from fits to Eqs. 9,11,13 (raw data reported by Stromberg [39]).

Sample Nb-49 had cylindrical shape with the length of about $2b$ = 12.7 mm and the diameter of $2a$ = 0.635 mm (i.e. $\frac{b}{a} = 20$) and, thus, by utilizing Eqs. 5,6 (proposed by Brandt [23]), one can obtain:

$$B_{MO} = B_{c1} = 1.03 \times B_p^{disk} \cong B_p^{disk} \quad (17)$$

Based on this, deduced $B_p(T)$ represents the lower critical field (Figure 11 and Equations 2,3).

When $B_{c1}(T)$ data were deduced, many research groups [24,30,32,42,43] fitted the deduced dataset to a parabolic model to determine $\lambda(0)$:

$$B_{c1}(T) = B_{c1}(0) \times \left(1 - \left(\frac{T}{T_c}\right)^2\right) = \frac{\phi_0}{4\pi\lambda^2(0)} \times \left(\ln\left(\frac{\lambda(0)}{\xi(0)}\right)\right) \times \left(1 - \left(\frac{T}{T_c}\right)^2\right) \quad (18)$$

or

$$B_{c1}(T) = \frac{\phi_0}{4\pi\lambda^2(0)} \times \left(\ln\left(\frac{\lambda(0)}{\xi(0)}\right) + 0.5\right) \times \left(1 - \left(\frac{T}{T_c}\right)^2\right) \quad (19)$$

. If experimental capability allows the application of a reasonably large magnetic field, then the upper critical field data, $B_{c2}(T)$, is also measured [32,42]. By using several analytical approximative functions for the temperature-dependent $B_{c2}(T)$ [44], the ground-state superconducting coherence length, $\xi(0)$, can be deduced for given materials and can be substituted into Eqs. 18,19.

A deeper analysis of the deduced $B_{c1}(T)$ data can be based on mentioned above Eqs. 1-4, which shows that $B_{c1}(T)$ is directly linked to the bulk density of Cooper pairs in the material. Thus, the analysis of $B_{c1}(T)$ can reveal the primary superconducting parameters of the material, for instance, the ground state energy gap, $\Delta(0)$, and the gap-to-transition temperature ratio of $2\Delta(0)/k_BT_c$. Below, we show how this analysis can be performed for the deduced $B_{c1}(T)$ dataset for niobium (sample Nb-49 (Fig. 12)).

First, it should be mentioned that the Ginzburg-Landau parameter $\kappa(T) = \frac{\lambda(T)}{\xi(T)}$ is temperature-dependent. Although this dependence is practically flattened, because $\kappa(T)$ is under the logarithm in Eq. 3, it is not difficult to implement the temperature dependence of this parameter by utilizing the approximative function proposed by Gor'kov [45]:

$$\kappa\left(\frac{T}{T_c}\right) = \kappa(0) \times \left(1 - 0.243\left(\frac{T}{T_c}\right)^2 + 0.039\left(\frac{T}{T_c}\right)^4\right). \quad (20)$$

Second, the London penetration depth, $\lambda(T)$, for *s*-wave superconductors can be expressed as [46]:

$$\lambda(T) = \frac{\lambda(0)}{\sqrt{1 - \frac{1}{2k_BT}\int_0^\infty \frac{d\varepsilon}{cosh^2\left(\frac{\sqrt{\varepsilon^2+\Delta^2(T)}}{2k_BT}\right)}}} \quad (21)$$

where $k_B$ is the Boltzmann constant, and $\Delta(T)$ is the superconducting energy gap, for which in our previous works [36,47,48] we used expression given by Gross *et al* [49,50]:

$$\Delta(T) = \Delta(0) \times \tanh\left[\frac{\pi \cdot k_B \cdot T_c}{\Delta(0)} \times \sqrt{\eta \times \frac{\Delta C}{\gamma T_c} \times \left(\frac{T_c}{T} - 1\right)}\right] \quad (22)$$

where $\frac{\Delta C}{\gamma T_c}$ is the relative jump in the electronic specific heat at $T_c$ (where $\gamma$ is so-called Sommerfeld constant), $\eta$ = 2/3 for *s*-wave superconductors, and temperature-dependent lower critical field for s-wave superconductors is:

$$B_{MO}(T) = B_p(T) = B_{c1}(T) = \frac{\phi_0}{4\pi}\frac{ln\left(1+\sqrt{2}\kappa(0)\times\left(1-0.243\left(\frac{T}{T_c}\right)^2+0.039\left(\frac{T}{T_c}\right)^4\right)\right)}{\lambda^2(0)}\left[1 - \frac{1}{2k_BT}\int_0^\infty \frac{d\varepsilon}{cosh^2\left(\frac{\sqrt{\varepsilon^2+\Delta^2(T)}}{2k_BT}\right)}\right]. \quad (23)$$

In the result, five major superconducting parameters of the superconductor (i.e. the transition temperature, $T_c$, ground state London penetration depth, $\lambda(0)$, the relative jump in electronic specific heat at $T_c$, $\frac{\Delta C}{\gamma T_c}$, and ground state energy gap, Δ(0), and the gap-to-transition temperature ratio, $\frac{2\Delta(0)}{k_B T_c}$) can be deduced from the fit of $B_{c1}(T)$ to Eq. 23.

Since 1965, niobium was considering to be two-band superconductor [51,52] where the smaller gap starts to manifest at $T \lesssim \frac{T_c}{7}$. As it was mentioned by Gor'kov and Kresin [53], the most superconductors exhibit two superconducting bands. However, to observe the second band in experiment [53], studied crystals should suffice the condition of $l \gg \xi(T)$ (where $l$ is the mean free path) to prevent the inter-band scattering which smears two-band picture. Because in Type-I and low-κ superconductors this condition is difficult to suffice, the two-gap picture is washed out and the one-gap picture is observed in samples. In this regard, sample Nb-49 prepared and studied by Stromberg [39] satisfies the condition because this sample has a very high resistance ratio of $\frac{R(300\,K)}{R(4.2\,K)} = 1360$.

Truly, an upturn in $B_{c1}(T)$ at $T \lesssim 2.5\,K$ can be seen in Fig. 12, which is an evidence for the opening of the second superconducting band. Similar upturn in the self-field critical current density, $J_c(sf,T)$, which is given by equation [17,18,54]:

$$J_c(sf,T) = \frac{\phi_0}{4\pi\mu_0} \times \frac{ln\left(1+\sqrt{2}\cdot\frac{\lambda(T)}{\xi(T)}\right)}{\lambda^3(T)} = \frac{\phi_0}{2\sqrt{2}\pi}\frac{{\mu_0}^{1/2}\cdot e}{{m_e}^{1/2}} \times n^{3/2}(T) \times ln\left(1+\sqrt{2}\frac{\lambda(T)}{\xi(T)}\right) \qquad (24)$$

where $\mu_0$ is the permeability of free space. The upturn was observed in perfect niobium thin films by Talantsev *et al* [55]. To fit the total measured $J_c(sf,T)$ dataset for niobium, a two-band model was used [55]:

$$J_{c,total}(sf,T) = J_{c,band_1}(sf,T) + J_{c,band_2}(sf,T) \qquad (25)$$

Details can be found in Fig. 8 of Ref. 55. By analogue, here we fitted total measured $B_{c1}(T)$ dataset for niobium to two-band model:

$$B_{c1,total}(T) = B_{c1,band_1}(T) + B_{c2,band_1}(T) \qquad (26)$$

where the lower critical field for each band is described by Eq. 21. We also assumed that both bands have the same $\kappa(0) = 1.0$ [2,38,39], and based on the available $B_{c1,total}(T)$ dataset, we equalized the relative jump in electronic specific heat at $T_c$ for both bands, $\frac{\Delta C_1}{\gamma_1 T_{c,1}} = \frac{\Delta C_2}{\gamma_2 T_{c,2}}$. The reason for the latter originates from the fact that raw experimental dataset should have a sufficiently large experimental dataset which cover as wide as possible temperature range of $0\,K < T \leq T_c$. However, it can be seen in Fig. 12 that the raw $B_{c1,total}(T)$ dataset was measured down to the lowest temperature of $T = 1.154\,K$ and, thus, the smaller band with $T_{c,2} \sim 2.5\,K$, does not have wide temperature range to be fitted to Eq. 23 for all the four fitting parameters to be freely fitted. This problem was discussed in Refs. 53,54 where we proposed a possible solution for this problem which is to use the same $\frac{\Delta C}{\gamma T_c}$ values for both bands, that:

$$\frac{\Delta C_1}{\gamma_1 T_{c,1}} = \frac{\Delta C_2}{\gamma_2 T_{c,2}} \qquad (27)$$

whereas this joint parameter is free fitting. The result of the $B_{c1,total}(T)$ fit to Eqs. 23,26,27 is shown in Fig. 13.

Overall, the fit confirms that niobium is a moderately strong electron-phonon coupled superconductor, for which deduced $\frac{2\Delta(0)}{k_B T_c}$ ratios for both bands are in a good agreement with each other. The deduced parameters and independently reported values are listed in Table I. It should be noted that the ratio for the smaller band, $\frac{2\Delta_2(0)}{k_B T_{c,2}}$, has a large uncertainty, originating from the aforementioned issue of the absence of experimental data below $T = 1.154\,K$.

### 3.8. $B_{c1}(T)$ data analysis for $ThIr_3$

Now, we return to polycrystalline $ThIr_3$, for which raw $B_{en}(T)$ data were reported by Górnicka *et al* [32] in their Fig. 3,c. By utilizing the demagnetization factor $N = 0.55$, we calculated the raw $B_{c1}(T)$ dataset and fitted this dataset to Eq. 23 (Fig. 14) (for this fit we used $\kappa(0) = 38$ [32]). All deduced parameters (Table II) show that this superconductor can be classified as moderately strong-coupled.

It is important to discuss the value of the Pauli-Chandrasekhar-Clogston limiting field, $B_{PCC}(0)$, which can be calculated from the deduced $\Delta(0)$. Simple-minded form of the equation described this field is [59]:

$$B_{PCC}(0) = \frac{\Delta(0)}{\sqrt{2}\times\mu_B} = 1.84 \times T_c \quad (28)$$

where $\mu_B$ denotes the Bohr magneton. However, this simplistic form is based on two hidden, but important, assumptions:

1. the Lande $g$-factor is:

$$g = 2 \quad (29)$$

and

2. the material is weak-coupled superconductor:

$$\Delta(0) = 1.76 \times k_B T_c \quad (30)$$

In general, these two assumptions are incorrect, and the proper expression for the Pauli-Chandrasekhar-Clogston limiting field is [60]:

$$B_{PCC}(0) = \frac{\Delta(0)}{\sqrt{g}\times\mu_B} \quad (31)$$

Thus, if the material exhibits strong charge carrier scattering, then $g$ can be well below 2. I addition, the superconductor can be either strong coupled, either exhibits *d*- or *p*-wave gap symmetry (for which $\Delta(0) > 2 \times k_B T_c$ [49,50]). All these features increase the $B_{PCC}(0)$ above the value, which can be calculated from the right hand side of Eq. 28, which is based only on the measured $T_c$ value.

It should be noted that the charge carrier scattering is not small in $ThIr_3$, which can be proven if one looks at the temperature dependent resistance shown in Fig. 4 in Ref. 32. In addition, our analysis (Fig. 14, Table II) revealed that $ThIr_3$ is a moderately strongly coupled superconductor. Based on that, calculated value for $B_{PCC}(0) = 9.1 \pm 0.7\ T$ (Table II) (in the assumption of $g = 2$) is well above extrapolated value for the upper critical field, $B_{c2}(0) = 4.7 - 4.9\ T$, reported by Górnicka *et al* [32].

### 3.9 Chiral noncentrosymmetric $TaRh_2B_2$ and $NbRh_2B_2$

Recently, Carnicom *et al* [42] reported DC magnetization, $M(B_{appl})$, magnetoresistivity, $\rho(T,B)$, and specific-heat, $C_p(T,B_{appl})$, data for two superconductors, $TaRh_2B_2$ and $NbRh_2B_2$, which exhibit chiral noncentrosymmetric crystal structure. Here, we analysed $B_{c1}(T)$ data recalculated from the reported $B_p(T)$ datasets (Figures 2(c,d) of Ref. [42]). In these calculations we used the demagnetization factor, *N*, which was reported in Ref. 42 for both samples.

The fit of $B_{c1}(T)$ data for $TaRh_2B_2$ to Eq. 23 (for which we used $\kappa(0) = 48$ [42]) is shown in Fig. 15 and the deduced parameters are listed in Table III. Due to raw $B_{c1}(T)$ dataset has only nine data points, with the purpose to reduce the uncertainty for the deduced free-fitting parameters, we fixed $T_c$ value in Eq. 23 to the experimentally measured value of $T_c = 5.8\ K$ (Fig. 2(a) in Ref. 42). It can be seen (Fig. 15 and Table III) that there is excellent agreement between the values of $\frac{\Delta C}{C}$ and of $\frac{2\Delta(0)}{k_B T_c}$ (deduced from the specific-heat data) reported by Carnicom *et al* [42] and our values deduced from the analysis of $B_{c1}(T)$. Overall, our analysis confirmed the result of Carnicom *et al* [42] that $TaRh_2B_2$ is a moderately strong coupled superconductor.

In regard of the Pauli-Chandrasekhar-Clogston limiting field, $B_{PCC}(0) = 11.9 \pm 0.6\ T$, which was calculated based on deduced $\Delta(0) = 0.976 \pm 0.054\ meV$ and assumed $g = 2$. This value is well matched to the extrapolated ground state upper critical field $B_{c2}(0) = 11.7\ T$ reported by Carnicom *et al* [42]. However, it should be noted that charge carrier scattering is large in $TaRh_2B_2$, as can be seen in Fig. 3(a) of Ref. 40, where the temperature-dependent resistivity, $\rho(T,B_{appl} = 0)$, is practically constant within the full temperature range of 10 K < *T*

< 300 K. Based on this, it is likely that the Lande $g$-factor is $g < 2$ and, thus, $B_{c2}(0)$ does not exceed $B_{PCC}(0)$.

However, if one looks at $\rho(T,B_{appl} = 0)$ data for $NbRh_2B_2$ (Fig. 3,b [42]) compound, it can be seen that the charge carrier scattering is very large in this compound, because $\rho(T,B_{appl} = 0)$ increases while the sample is cooled down. Based on this, it is not surprizing that the calculated $B_{PCC}(0)$, assuming $g = 2$, is well below the extrapolated ground state upper critical field $B_{c2}(0) = 14.1\ T$ reported by Carnicom *et al* [42].

Additionally, the application of Eq. 23 to fit $B_{c1}(T)$ data for $NbRh_2B_2$ returned different free-fitting parameters from those ones reported by Carnicom *et al* [42] (see Figure 16 and Table IV). For the fit we used $\kappa(0) = 51$ [42]). In particular, the deduced ground state energy gap is:

$$\Delta(0) = 1.13 \pm 0.04\ meV, \tag{32}$$

and the ratio:

$$\frac{2\Delta(0)}{k_B T_c} = 3.45 \pm 0.12, \tag{33}$$

which indicates that this compound is weak-coupled superconductor.

Because all deduced values were significantly different from the values reported by Carnicom *et al* [42] (see, Table IV), we performed an additional test. To disprove/reaffirm the values deduced by our analysis, we performed a fit of the temperature-dependent electronic specific heat $C_{el}(T,B)$ (reported in Fig. 2,f in Ref. 42) to the low-$T$ analytical asymptote:

$$C_{el}(T, B = 0) = \gamma_0 T + A \times e^{-\frac{\Delta(0)}{k_B T}} \tag{34}$$

where $\gamma_0$, $A$, and $\Delta(0)$ are the free-fitting parameters. It should be stressed that all low-$T$ asymptotes of the Bardeen-Cooper-Schrieffer theory, which for $s$-wave superconductors were first proposed by Mühlschlegel [61], are accurate approximations at:

$$\frac{T}{T_c} \lesssim \frac{1}{3} \tag{35}$$

Thus, we limited $C_{el}(T, B = 0)$ to the minimal dataset, for which the fit to Eq. 32 can be converged. This value was calculated for:

$$T \leq 3.4\, K \tag{36}$$

and the fit is shown in Fig. 16(b). It can be observed in Fig. 16(b), that the deduced

$$\Delta(0) = 1.14 \pm 0.04\, meV \tag{37}$$

is in unprecedent agreement with the value of $\Delta(0)$ deduced from the analysis of $B_{c1}(T)$ (Fig. 16(a), Eq. 32). Calculated

$$B_{PCC}(0)_{g=2} = 13.8 \pm 0.5\, T \tag{38}$$

is different, but not overwhelmingly different with the extrapolated $B_{c2}(0) = 18.0\, T$ [42]. We can estimate the Lande $g$-factor for $NbRh_2B_2$ by equalizing $B_{PCC}(0)$ and $B_{c2}(0)$, from which:

$$g(NbRh_2B_2) \cong 1.2 \tag{39}$$

## 5. Conclusions

In conclusion, in this paper, we propose a mathematical routine to deduce the temperature dependent first flux entry field data, $B_p(T)$, from DC magnetization data, $M(B_{appl})$, by utilizing a power-law fitting function for which the threshold criterion $M_c$ can be chosen based on convention, similar to the electric field criterion of $E_c = 1$ μV/cm, which is widely used to define critical currents in superconductors. The routine can be an additional and, from our point of view, a more formalized mathematical tool to extract $B_p(T)$ data from experimental DC magnetization curves.

The proposed routine, from other hand, has reasonably strict requirements for the raw experimental data. One of the most important requirements is that measurements should be performed by using a very small steps of $B_{appl}$, especially for samples with the power-law exponent $n \lesssim 5$ to deduce $B_p(T)$ dataset with reasonable accuracy.

In addition, we demonstrated how deduced/reported $B_p(T)$ datasets can be analysed to extract the ground state energy gap, $\Delta(0)$, gap-to-transition temperature ratio, $\frac{2\Delta(0)}{k_B T_c}$, and specific heat jump at the transition temperature at $T_c$, $\frac{\Delta C}{C}$, for s-wave superconductors.

## Data Availability Statement

The data that support the findings of this study are available from the author upon reasonable request.

## Conflicts of Interest

The authors have no conflicts to disclose.

## Acknowledgements

The author acknowledges financial support provided by the Ministry of Science and Higher Education of Russia (theme “Pressure” No. AAAA-A18-118020190104-3).

## Figures Captions

**Figure 1.** Full scale DC $M(B_{appl}, T = 1.7\ K)$ dataset for polycrystalline $ThIr_3$ (raw data reported by Górnicka *et al* [32]) where used for the fit raw $M(B_{appl}, T = 1.7\ K)$ data is shown by orange balls. Based on observed $|M(B_{appl}, T = 1.7\ K)|_{max} = 13\ kA/m$ , we use $M_c = 0.02 \times |M(B_{appl}, T = 1.7\ K)|_{max} \cong 0.26\ \frac{kA}{m}$ and $M(B_{appl}) - M_0 - k \times B_{appl} \lesssim 5.5 \times M_c$. Green ball shows deduced $M(B_p, T = 1.7\ K)$. Deduced values are $B_p = (1.9 \pm 0.1)\ mT$, $n = 2.27 \pm 0.09$ and $k = -0.194 \pm 0.002\ kA/(m \times mT)$.

**Figure 2.** The evolution of deduced $B_p$ (panels (a) and (d)), *n* (panels (b) and (e) and *k* (panels (c) and (f)) vs $B_{appl}/B_{M,max}$ (panels (a)-(c)) and $((M(B_{appl}) - k \times B_{appl})/M_c)$ (panels (d)-(f)) for the fits for polycrystalline $ThIr_3$. Raw $M(B_{appl}, T = 1.7\ K)$ data for polycrystalline $ThIr_3$ reported by Górnicka *et al* [32]. Ovals indicate the same group of deduced parameters for low $B_{appl}/B_{M,max}$ values. Green balls show the chosen values for which satisfy the condition of $\frac{M(B_{appl}) - M_0 - k \times B_{appl}}{M_c} \cong 5$.

**Figure 3.** $M_V(B_{appl}, T = 0.5\ K)$ data for $WB_{4.2}$ (raw data reported by Carnicom *et al* [24]) and data fit to Eqs. 9,11,13 with utilized $M_c = 0.02 \times |M(B_{appl}, T = 0.5\ K)|_{max} \cong 70\ \frac{A}{m}$. Deduced $M_0 = -30 \pm 20\ A/m$, $k = -180 \pm 4\ A/(m \times mT)$, $n = 3.4 \pm 1.1$, $B_p = 1.2 \pm 0.2\ mT$. $M_V(B_p, T = 0.5\ K)$ is shown by green ball, goodness of fit *R* = 0.9998.

**Figure 4.** The evolution of deduced $B_p$ (panels (a) and (d)), *n* (panels (b) and (e)) and *k* (panels (c) and (f)) vs $B_{appl}/B_{M,max}$ (panels (a)-(c)) and $((M(B_{appl}) - k \times B_{appl})/M_c)$ (panels (d)-(f)) for the fits for polycrystalline $WB_{4.2}$. Raw $M_V(B_{appl}, T = 0.5\ K)$ data for polycrystalline $WB_{4.2}$ reported by Carnicom *et al* [24]. Green balls show the chosen values for which satisfy the condition of $\frac{M(B_{appl}) - M_0 - k \times B_{appl}}{M_c} \cong 8$.

**Figure 5.** $M(B_{appl}, T = 1.85\ K)$ data for polycrystalline $BaTi_2Bi_2O$ (raw data reported by Yajima [33]) and data fit to Eqs. 9,11,13. The fits were performed for: (a) $B_{appl} < 5.5\ mT$, (b) $B_{appl} < 8\ mT$, (c) $B_{appl} \leq 12\ mT$, (d) $B_{appl} \leq 14\ mT$Deduced $M_0 = 0.017 \pm 0.001\ A \times m^2/kg$ $k = -0.052 \pm 0.001\ A \times m^2/kg \times mT$, $n = 2.13 \pm 0.06$, $B_p = 2.7 \pm 0.2\ mT$, $M_c = 0.02 \times |M(B_{appl}, T = 1.85\ K)|_{max} = 0.009\ A \times m^2/kg$, goodness of fit for all panels is better than *R* = 0.9997. 95% confidence bands are shown by pink shaded areas.

**Figure 6.** The evolution of deduced $B_p$ (panels (a) and (d)), *n* (panels (b) and (e)) and *k* (panels (c) and (f)) vs $B_{appl}/B_{M,max}$ (panels (a)-(c)) and $((M(B_{appl}) - k \times B_{appl})/M_c)$ (panels (d)-(f)) for the fits for polycrystalline $BaTi_2Bi_2O$ (raw data reported by Yajima [33]) and data fit to Eqs. 9,11,13. Green balls show the chosen values for which satisfy the condition of $\frac{M(B_{appl}) - M_0 - k \times B_{appl}}{M_c} \cong 31$.

**Figure 7.** $M(B_{appl}, T = 2.0\ K)$ data for $Th_4H_{15}$ (raw data reported by Wang *et al* [34]) and data fit to Eqs. 9,11,13. Deduced $M_0 = 0.005 \pm 0.022\ A \times m^2/kg$, $k = -0.193 \pm 0.009\ A \times m^2/kg \times mT$, $n = 2.4 \pm 0.2$, $B_p = 4.9 \pm 0.8\ mT$, $M_c = 0.05\ A \times m^2/kg$, goodness of fit *R* = 0.9991. 95% confidence bands are shown by pink shaded areas.

**Figure 8.** $M\left(B_{appl}, T = 2.0\ K\right)$ data for lead thin films deposited on mica substrate (raw data reported by Lock [35]) and data fits to Eqs. 9,11,13. (a) film thickness is 760 nm, (b) film thickness is 334 nm, (c) film thickness is 181 nm, (d) film thickness is 107 nm. Deduced values are shown in panels. In panel ***a*** deduced $B_p$ is shown by arrow due to deduced value overlaps with the datapoint. Goodness of fit for panels ***a-c*** are $R > 0.9990$, and goodness of fit for panel ***d*** is $R = 0.994$. 95% confidence bands are shown by pink shaded areas.

**Figure 9.** Deduced $B_p$ and $n$ values vs thickness, $2b$, for lead films.

**Figure 10.** $M\left(B_{appl}, T = 5.0\ K\right)$ data for epitaxial $MgB_2$ thin films (raw data reported by Tan *et al.* [37]) and data fit to Eqs. 9,11 with utilized $M_c = 0.005 \times \left|M\left(B_{appl}, T = 5.0\ K\right)\right|_{max} \cong 0.004\ \frac{A}{m}$. Deduced $M_0 = -12 \pm 2\ mA/m$, $k = -4.95 \pm 0.06\ A/(m \times T)$, $n = 3.3 \pm 0.3$, $B_p = 61 \pm 6\ mT$. $M(B_{en}, T = 5.0\ K)$ is shown by green ball, goodness of fit $R = 0.9997$.

**Figure 11.** $M\left(B_{appl}, T\right)$ data for Nb single crystal (sample Nb-49 [39]) at several temperatures: (a) $T = 7.966$ K, (b) $T = 7.715$ K, (c) $T = 3.985$ K, (d) $T = 3.700$ K. Raw $M(B_{appl})$ data reported by Stromberg [39]). Fits to Eqs. 9,11,13 and deduced values are shown in panels. Goodness of fit for all panels $R > 0.9997$. 95% confidence bands are shown by pink shaded areas.

**Figure 12.** Temperature dependent $B_p(T)$ and $n(T)$ deduced for single crystal of niobium (sample Nb-49) from fits to Eqs. 9,11,13 (raw data reported by Stromberg [39]).

**Figure 13.** $B_{c1}(T)$ data and fit to Eqs. 23,26,27 for pure niobium single crystal (sample Nb-49 for which raw $M\left(B_{appl}, T\right)$ reported by Stromberg [39]). Subscript indexes 1 and 2 designate superconducting Band 1 and Band 2, respectively. Deduced parameters are $\lambda_{total}(0) = 29.7 \pm 1.1\ nm$, $\frac{\Delta C_1}{\gamma_1 T_{c,1}} = \frac{\Delta C_2}{\gamma_2 T_{c,2}} = 2.1 \pm 0.2$; Band 1: $T_{c,1} = 9.25 \pm 0.05\ K$, $\Delta_1(0) = 1.71 \pm 0.06\ meV$, $\frac{2\Delta_1(0)}{k_B T_{c,1}} = 4.3 \pm 0.2$, $\lambda_1(0) = 31.1 \pm 0.1\ nm$; Band 2: $T_{c,2} = 2.6 \pm 0.1\ K$, $\Delta_2(0) = 0.48 \pm 0.33\ meV$, $\frac{2\Delta_2(0)}{k_B T_{c,2}} = 4.2 \pm 3.0$, $\lambda_1(0) = 99 \pm 13\ nm$. Goodness of fit is 0.9986. 95% confidence bands are shown by pink area.

**Figure 14.** $B_{c1}(T)$ data and fit to Eq. 21 for polycrystalline $ThIr_3$ sample for which raw $B_{en}(T)$ data was reported by Górnicka *et al* [32]). Utilized $\kappa(0) = 38$ [32]. Deduced parameters are: $T_c = 4.6 \pm 0.1\ K$, $\lambda(0) = 349 \pm 5\ nm$, $\frac{\Delta C}{\gamma T_c} = 2.2 \pm 0.5$, $\Delta(0) = 747 \pm 59\ \mu eV$, $\frac{2\Delta(0)}{k_B T_c} = 3.8 \pm 0.3$ Goodness of fit is 0.9957. 95% confidence bands are shown by pink and green areas.

**Figure 15.** $B_{c1}(T)$ data and fit to Eq. 23 for polycrystalline $TaRh_2B_2$ sample for which raw $B_p(T)$ data was reported by Carnicom *et al* [42]. Utilized $\kappa(0) = 48$ [42] and $T_c = 5.8\ K$ (fixed). Deduced parameters are: $\lambda(0) = 285 \pm 2\ nm$, $\frac{\Delta C}{\gamma T_c} = 1.7 \pm 0.1$, $\Delta(0) = 976 \pm 54\ \mu eV$, $\frac{2\Delta(0)}{k_B T_c} = 3.9 \pm 0.2$. Goodness of fit is 0.9985. 95% confidence bands are shown by pink and green areas.

**Figure 16.** $B_{c1}(T)$ (a) and $C_{el}(T)$ (b) data and fits for polycrystalline $TaRh_2B_2$ sample for which experimental data was reported by Carnicom *et al* [42]. (a) $B_{c1}(T)$ data and data fit to Eq. 23. Utilized $\kappa(0) = 51$ [42] and $T_c = 5.8\ K$ (fixed). Deduced parameters are: $\lambda(0) = 237 \pm 1\ nm$, $\frac{\Delta C}{\gamma T_c} = 1.7 \pm 0.1$, $\Delta(0) = 1.13 \pm 0.04\ meV$, $\frac{2\Delta(0)}{k_B T_c} = 3.45 \pm 0.12$. Goodness of fit is 0.9984. 95% confidence bands are shown by pink and green areas. (b) $C_{el}(T)$ data and data fit to Eq. 34, $\Delta(0) = 1.14 \pm 0.03\ meV$, $\frac{2\Delta(0)}{k_B T_c} = 3.48 \pm 0.09$. Goodness of fit is 0.9998. 95% confidence bands are shown by blue area.

# Tables

**Table I.** Deduced parameters for niobium single crystal (raw DC magnetization data reported by Stromberg [39], Sample Nb-49).

| **Parameter** | **Independently reported** | **Deduced, Band 1** | **Deduced, Band 2** |
|---|---|---|---|
| Energy gap, $\Delta(0)$ $(meV)$ | 1.55 [57,58] | $1.71 \pm 0.06$ | $0.48 \pm 0.33$ |
| London penetration depth, $\lambda(0)$ $(nm)$ | 31 [39] | $29.7 \pm 1.1$<br>$31.1 \pm 0.1$ | <br>$99 \pm 13$ |
| Relative jump in electronic specific heat at $T_c$, $\frac{\Delta C}{\gamma T_c}$ | 1.93 [58] | $2.1 \pm 0.2$ | |
| Gap-to-transition temperature ratio, $\frac{2\Delta(0)}{k_B T_c}$ | 3.9 [57,58] | $4.3 \pm 0.2$ | $4.2 \pm 3.0$ |

**Table II.** Deduced parameters for $ThIr_3$ polycrystal (raw $B_{c1}(T)$ data reported by Górnicka *et al* [32]).

| **Parameter** | **BCS weak-coupling limit** | **Reported by Górnicka *et al* [32]** | **Deduced from $B_{c1}(T)$** |
|---|---|---|---|
| Lower critical field, $B_{c1}(0)$ $(mT)$ | | 6.0 | $5.4 \pm 0.3$ |
| London penetration depth, $\lambda(0)$ $(nm)$ | | 315 | $349 \pm 5$ |
| Energy gap, $\Delta(0)$ $(\mu eV)$ | | 500 | $747 \pm 59$ |
| Transition temperature, $T_c$ (K) | | 4.41 | $4.6 \pm 0.1$ |
| Relative jump in electronic specific heat at $T_c$, $\frac{\Delta C}{\gamma T_c}$ | 1.43 | 1.6 | $2.2 \pm 0.5$ |
| Gap-to-transition temperature ratio, $\frac{2\Delta(0)}{k_B T_c}$ | 3.53 | 2.6 | $3.8 \pm 0.3$ |
| Upper critical field, $B_{c2}(0)$ (T) | | 4.7-4.9 | |
| Pauli-Chandrasekhar-Clogston limiting field, $B_{PCC}(0)$ (T) ($g$ = 2) | | 6.1 (based on $\Delta(0) = 0.5$ $meV$) | $9.1 \pm 0.7$ |

**Table III.** Deduced parameters for $TaRh_2B_2$ polycrystal (raw $B_{c1}(T)$ data reported by Carnicom *et al* [42]).

| **Parameter** | **BCS weak-coupling limit** | **Reported by Carnicom *et al* [42]** | **Deduced from $B_{c1}(T)$** |
|---|---|---|---|
| Lower critical field, $B_{c1}(0)$ $(mT)$ | | 9.56 | $8.6 \pm 0.3$ |
| London penetration depth, $\lambda(0)$ $(nm)$ | | 258 | $285 \pm 2$ |
| Energy gap, $\Delta(0)$ $(meV)$ | | 0.98 | $0.976 \pm 0.054$ |
| Transition temperature, $T_c$ (K) | | 5.8 | 5.8 (fixed) |
| Relative jump in electronic specific heat at $T_c$, $\frac{\Delta C}{\gamma T_c}$ | 1.43 | 1.56 | $1.7 \pm 0.1$ |
| Gap-to-transition temperature ratio, $\frac{2\Delta(0)}{k_B T_c}$ | 3.53 | 3.9 | $3.9 \pm 0.2$ |
| Upper critical field, $B_{c2}(0)$ (T) | | 11.7 | |
| Pauli-Chandrasekhar-Clogston limiting field, $B_{PCC}(0)$ (T) ($g$ = 2) | | 10.7 | $11.9 \pm 0.6$ |

**Table IV.** Deduced parameters for $NbRh_2B_2$ polycrystal (raw $B_{c1}(T)$ data reported by Carnicom *et al* [42]).

| **Parameter** | **BCS weak-coupling limit** | **Reported by Carnicom *et al* [42]** | **Deduced from $B_{c1}(T)$** |
|---|---|---|---|
| Lower critical field, $B_{c1}(0)$ $(mT)$ | | 13.5 | $8.6 \pm 0.3$ |
| London penetration depth, $\lambda(0)$ $(nm)$ | | 219 | $285 \pm 2$ |
| Energy gap, $\Delta(0)$ $(meV)$ | | 1.4 | $1.13 \pm 0.04$ |
| Transition temperature, $T_c$ (K) | | 7.6 | 7.6 (fixed) |
| Relative jump in electronic specific heat at $T_c$, $\frac{\Delta C}{\gamma T_c}$ | 1.43 | 1.6 | $1.7 \pm 0.1$ |
| Gap-to-transition temperature ratio, $\frac{2\Delta(0)}{k_B T_c}$ | 3.53 | 4.3 | $3.45 \pm 0.12$ |
| Upper critical field, $B_{c2}(0)$ (T) | | 18.0 | |
| Pauli-Chandrasekhar-Clogston limiting field, $B_{PCC}(0)$ (T) ($g$ = 2) | | 14.1 | $13.8 \pm 0.5$ |

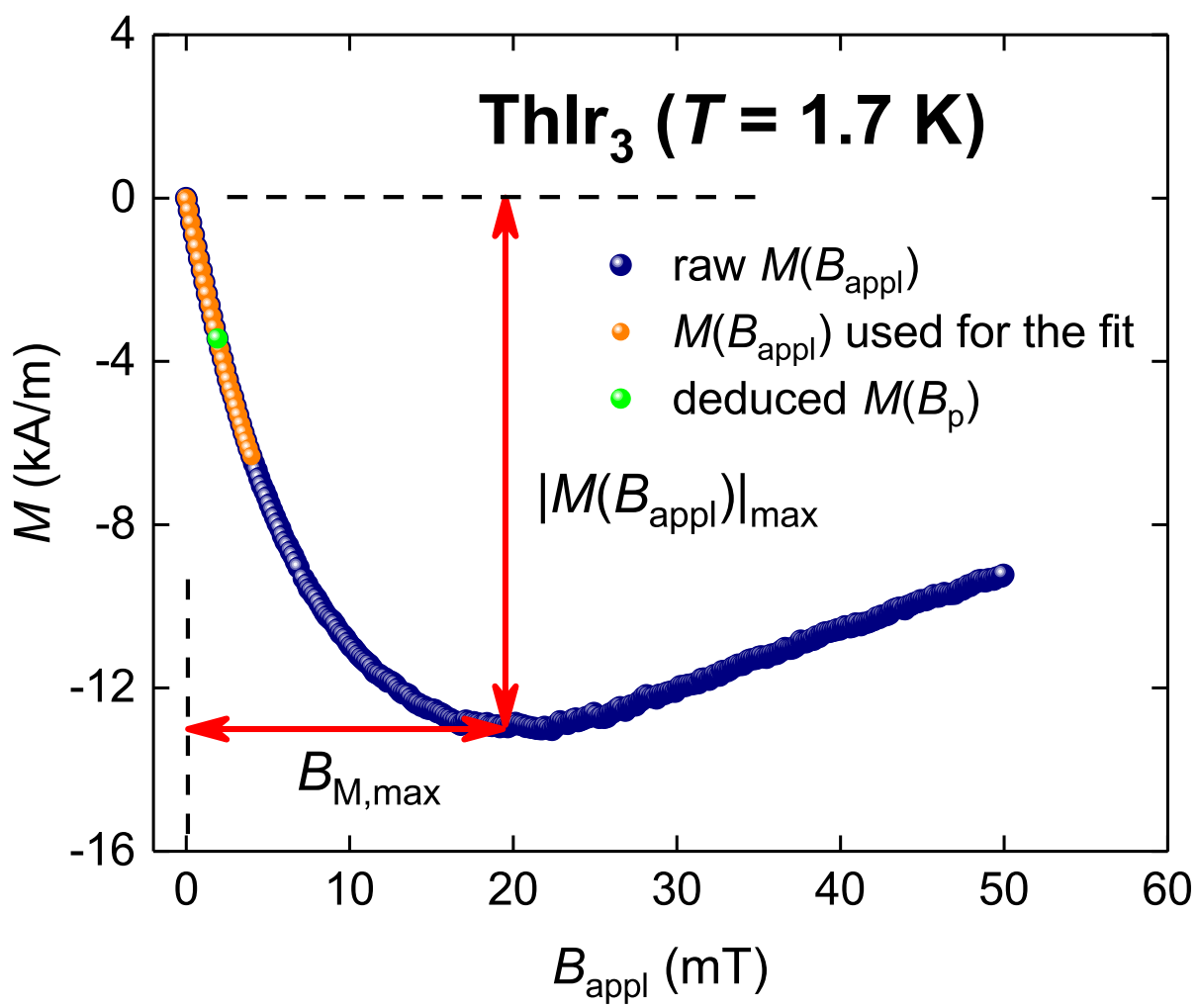

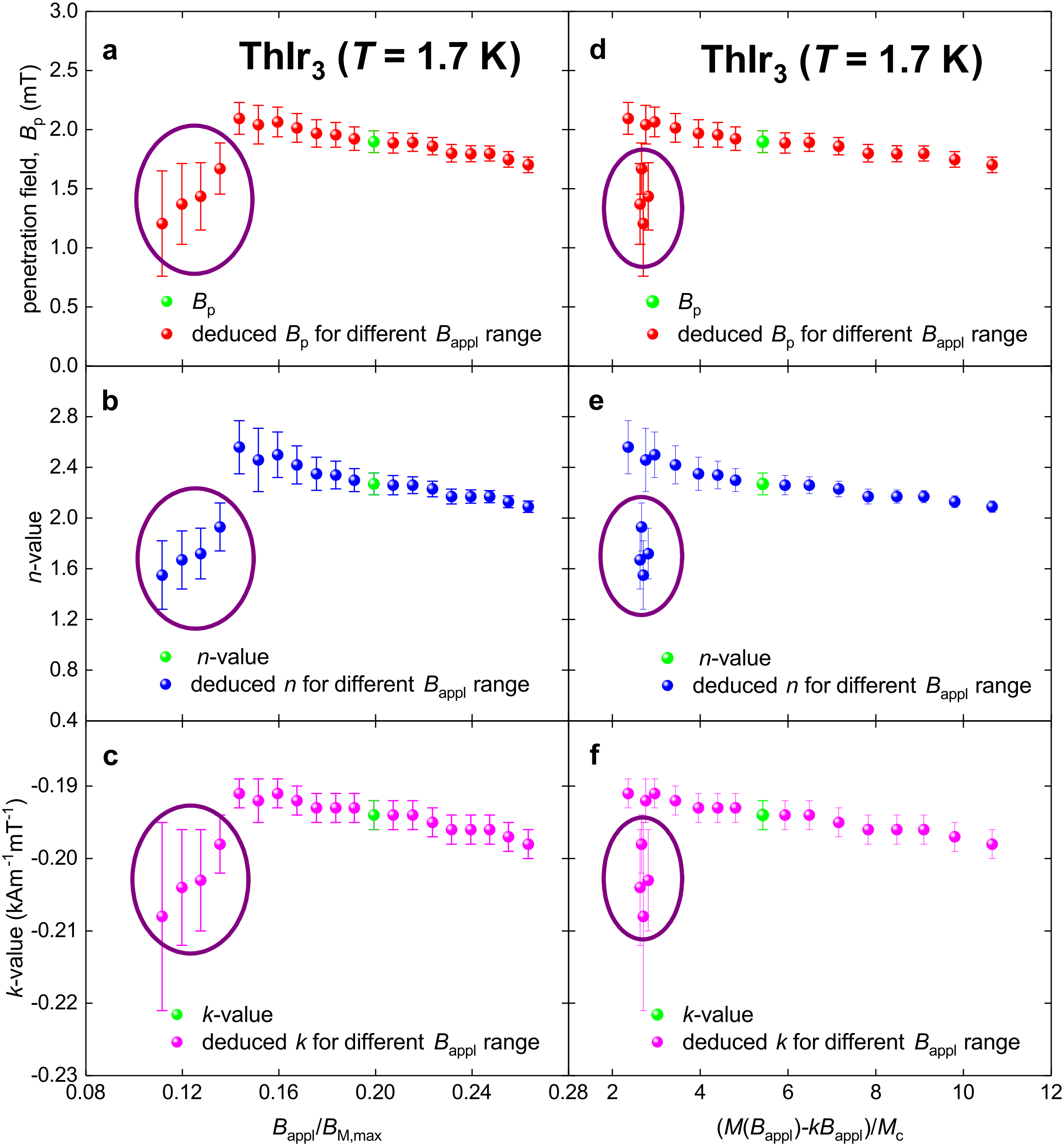

ThIr$_3$ ($T$ = 1.7 K)
a
b
c
d
e
f
penetration field, $B_p$ (mT)
$n$-value
$k$-value (kAm$^{-1}$mT$^{-1}$)
$B_p$
deduced $B_p$ for different $B_{appl}$ range
$n$-value
deduced $n$ for different $B_{appl}$ range
$k$-value
deduced $k$ for different $B_{appl}$ range
$B_{appl}/B_{M,max}$
$(M(B_{appl})-kB_{appl})/M_c$

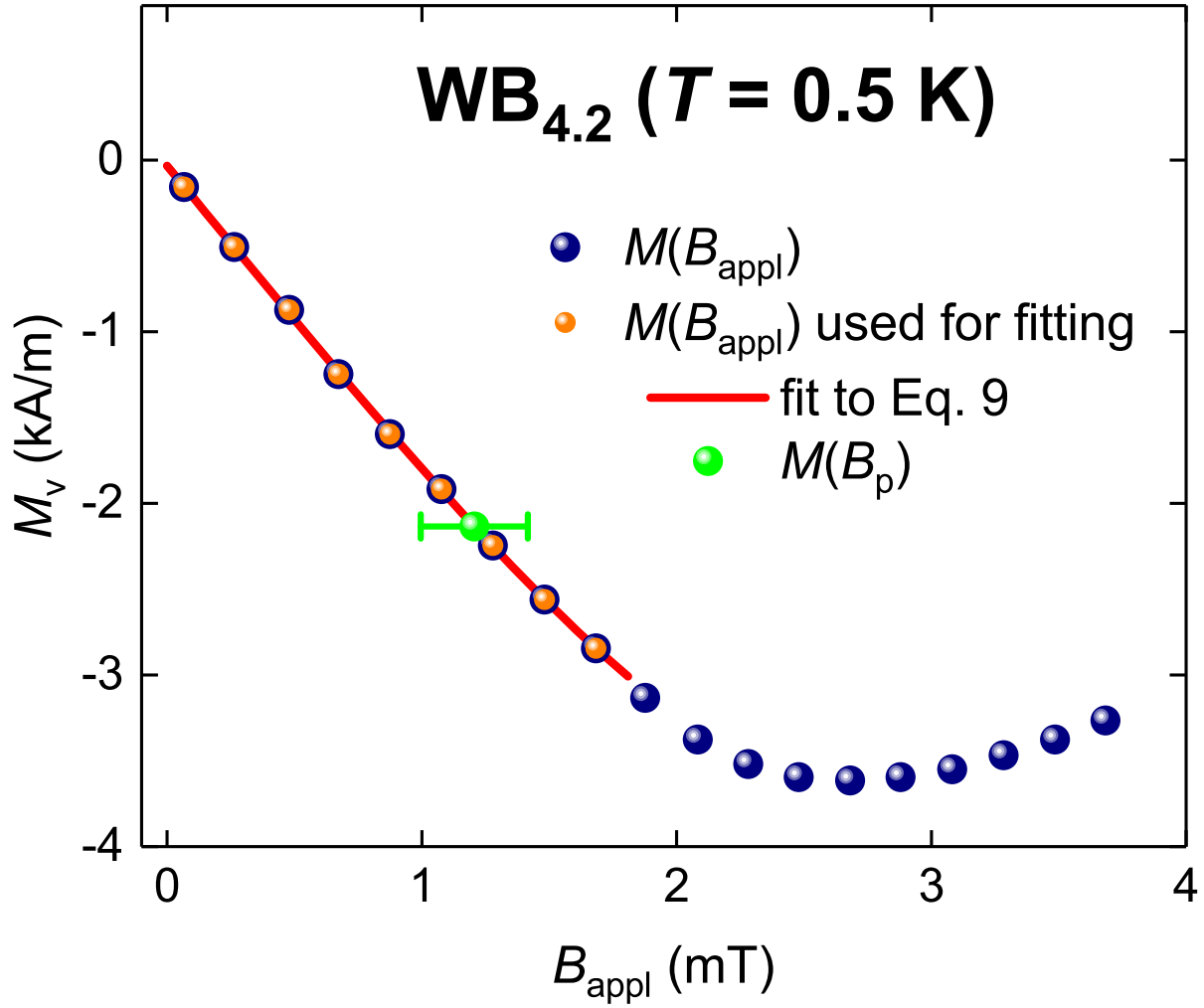

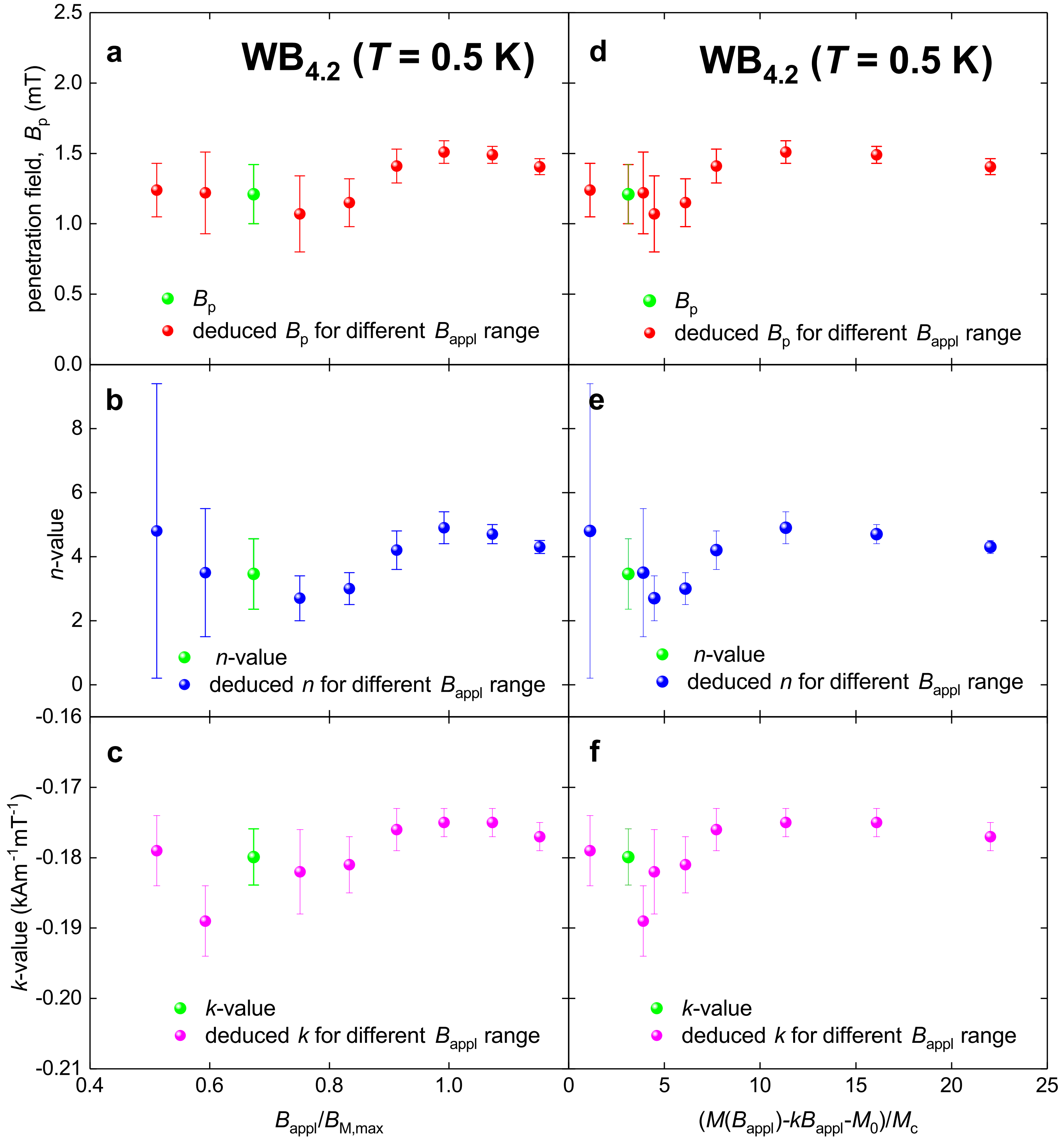

a
WB4.2 (T = 0.5 K)
penetration field, Bp (mT)
Bp
deduced Bp for different Bappl range
b
n-value
n-value
deduced n for different Bappl range
c
k-value (kAm-1mT-1)
k-value
deduced k for different Bappl range
Bappl/BM,max
d
WB4.2 (T = 0.5 K)
Bp
deduced Bp for different Bappl range
e
n-value
deduced n for different Bappl range
f
k-value
deduced k for different Bappl range
(M(Bappl)-kBappl-M0)/Mc

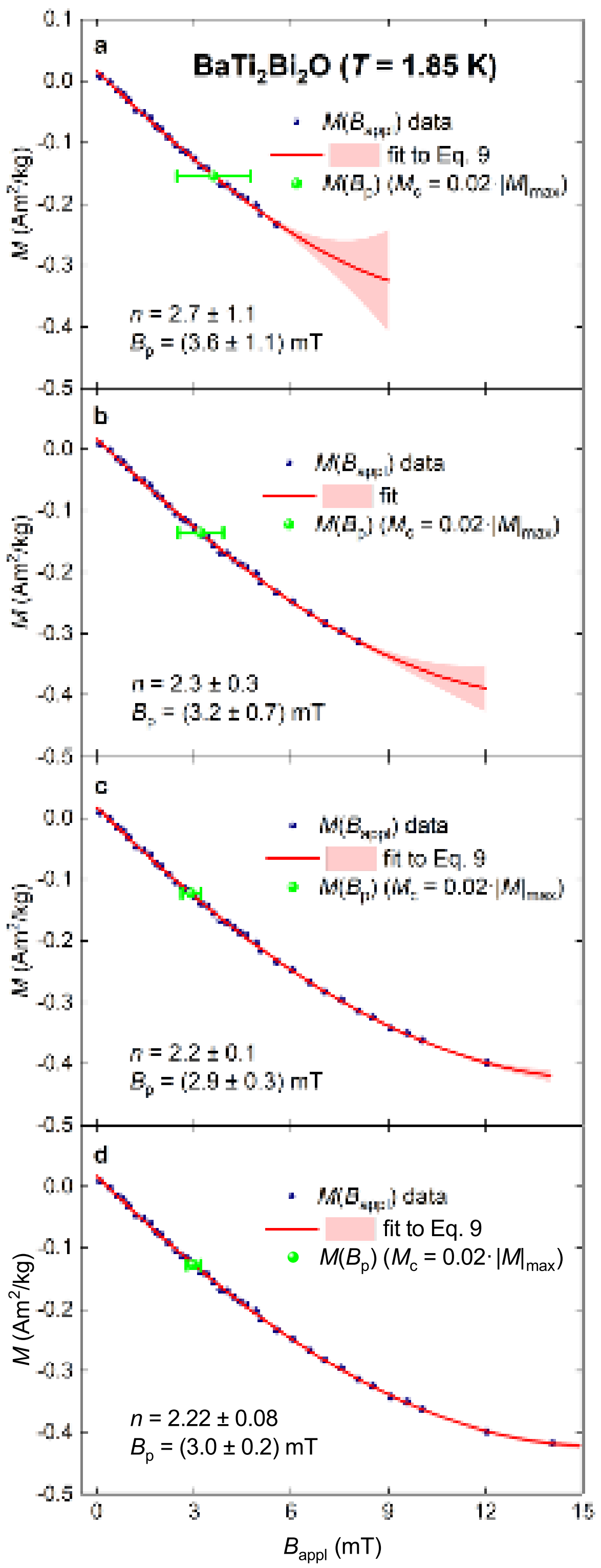

a
BaTi2Bi2O (T = 1.85 K)
M(Bappl) data
fit to Eq. 9
M(Bp) (Mc = 0.02·|M|max)
n = 2.7 ± 1.1
Bp = (3.6 ± 1.1) mT
b
M(Bappl) data
fit
M(Bp) (Mc = 0.02·|M|max)
n = 2.3 ± 0.3
Bp = (3.2 ± 0.7) mT
c
M(Bappl) data
fit to Eq. 9
M(Bp) (Mc = 0.02·|M|max)
n = 2.2 ± 0.1
Bp = (2.9 ± 0.3) mT
d
M(Bappl) data
fit to Eq. 9
M(Bp) (Mc = 0.02·|M|max)
n = 2.22 ± 0.08
Bp = (3.0 ± 0.2) mT
M (Am2/kg)
Bappl (mT)

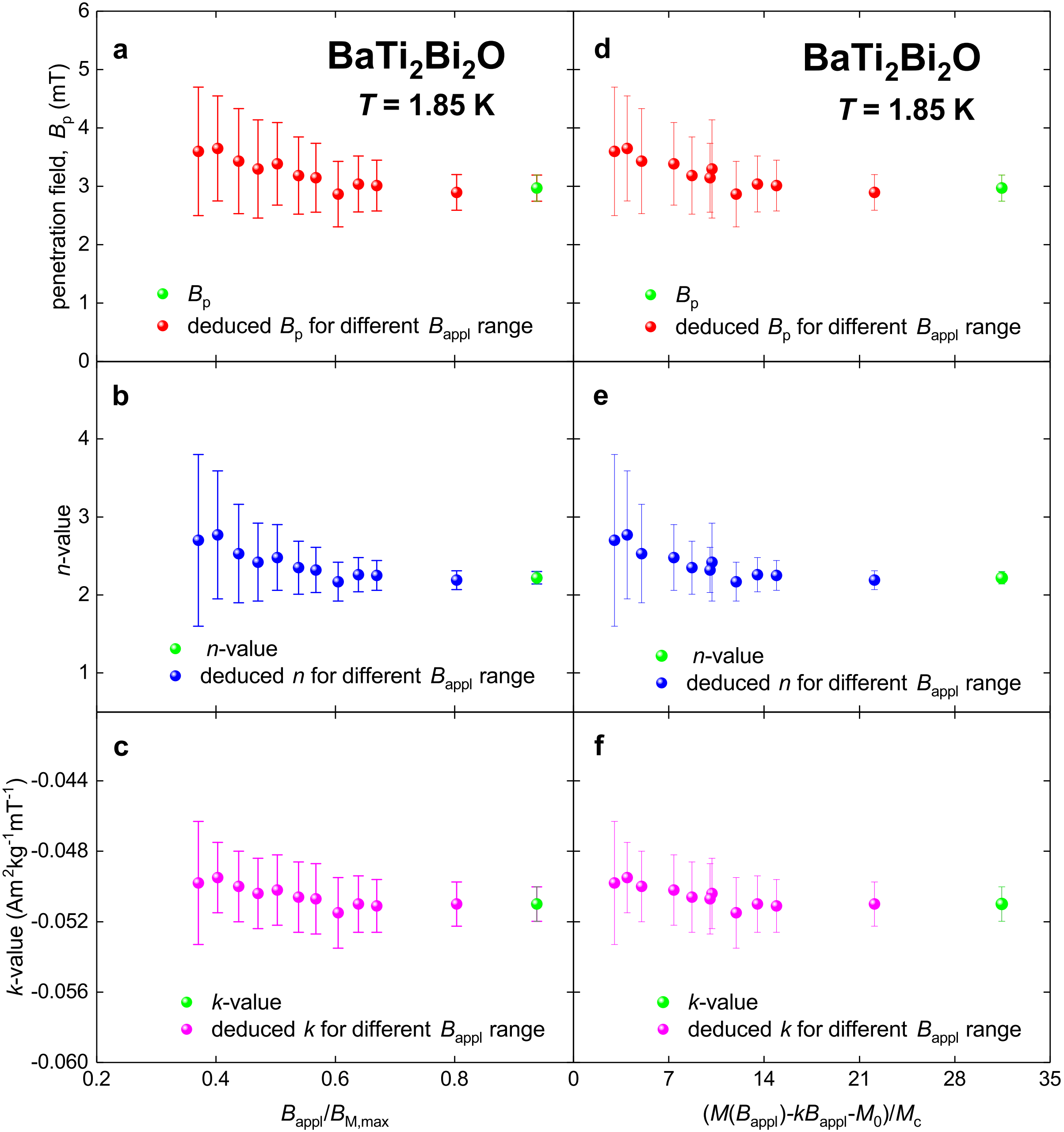

a
d
b
e
c
f
BaTi2Bi2O
T = 1.85 K
penetration field, Bp (mT)
n-value
k-value (Am2kg-1mT-1)
Bp
deduced Bp for different Bappl range
n-value
deduced n for different Bappl range
k-value
deduced k for different Bappl range
Bappl/BM,max
(M(Bappl)-kBappl-M0)/Mc

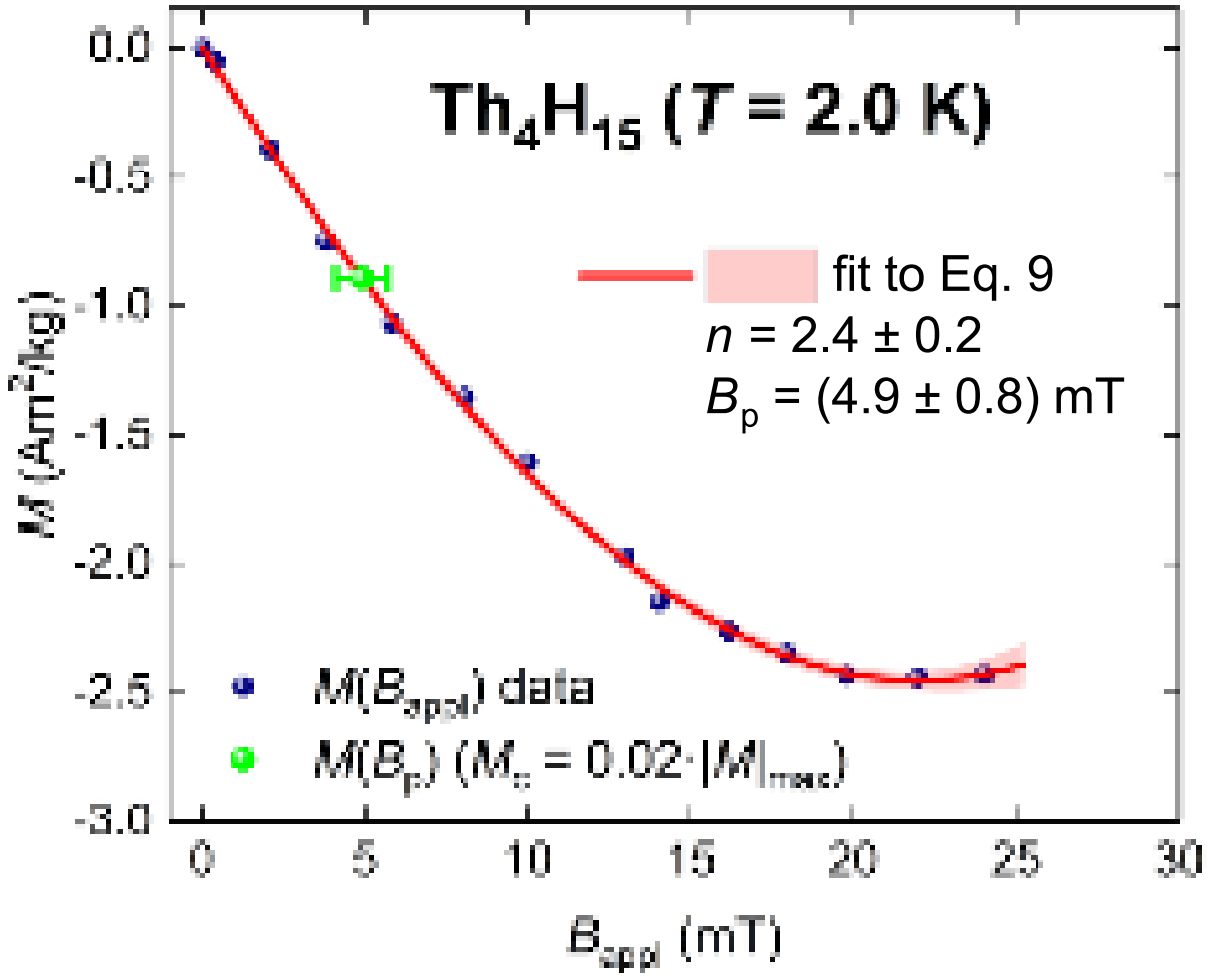

Th4H15 (T = 2.0 K)
fit to Eq. 9
n = 2.4 ± 0.2
Bp = (4.9 ± 0.8) mT
M(Bappl) data
M(Bp) (Mc = 0.02·|M|max)
M (Am²/kg)
Bappl (mT)
0.0
-0.5
-1.0
-1.5
-2.0
-2.5
-3.0
0
5
10
15
20
25
30

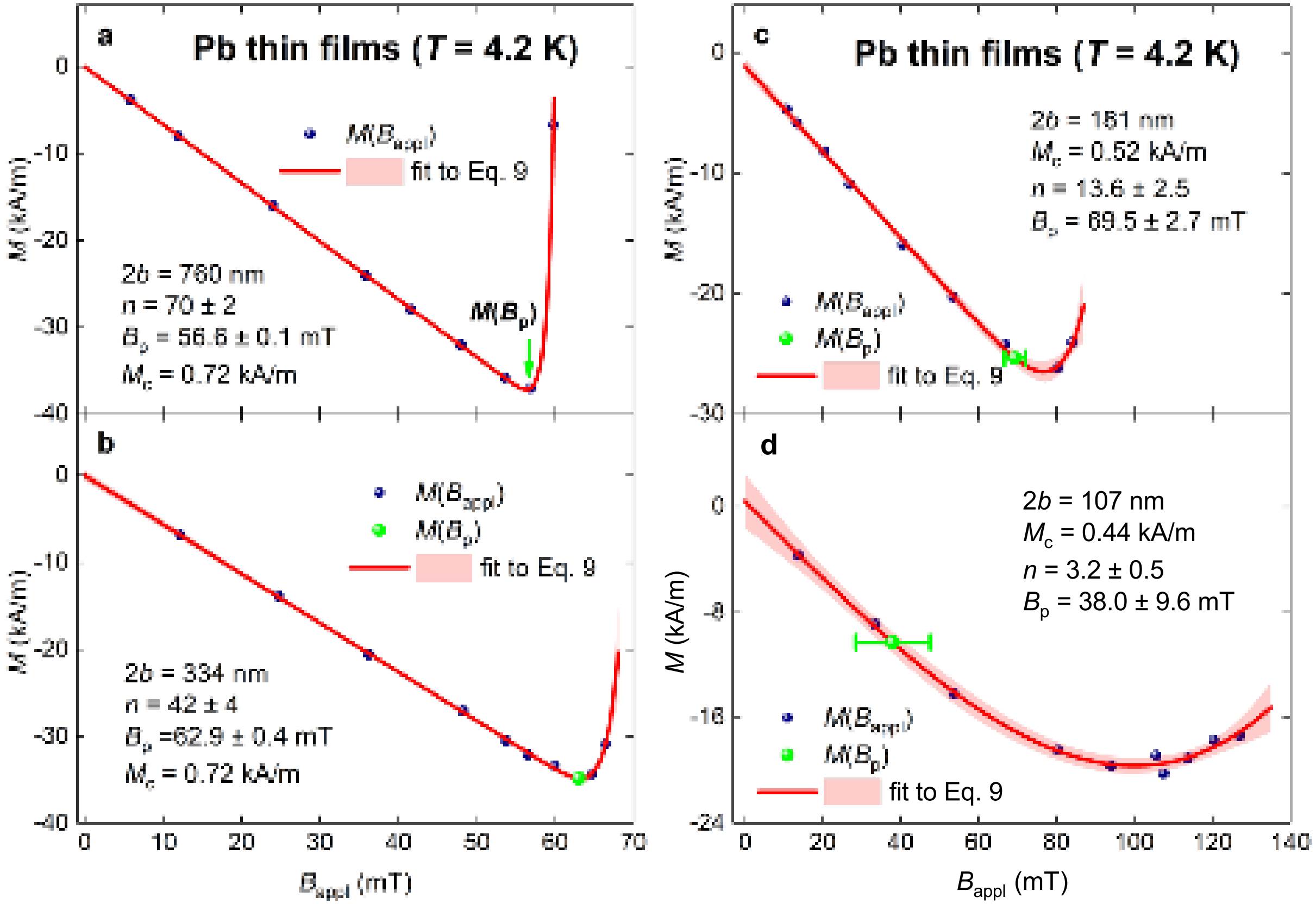

a
Pb thin films (T = 4.2 K)
M(B_appl)
fit to Eq. 9
2b = 760 nm
n = 70 ± 2
B_p = 56.6 ± 0.1 mT
M_c = 0.72 kA/m
M(B_p)
M (kA/m)
b
M(B_appl)
M(B_p)
fit to Eq. 9
2b = 334 nm
n = 42 ± 4
B_p =62.9 ± 0.4 mT
M_c = 0.72 kA/m
M (kA/m)
B_appl (mT)
c
Pb thin films (T = 4.2 K)
2b = 161 nm
M_c = 0.52 kA/m
n = 13.6 ± 2.5
B_p = 69.5 ± 2.7 mT
M(B_appl)
M(B_p)
fit to Eq. 9
M (kA/m)
d
2b = 107 nm
M_c = 0.44 kA/m
n = 3.2 ± 0.5
B_p = 38.0 ± 9.6 mT
M(B_appl)
M(B_p)
fit to Eq. 9
M (kA/m)
B_appl (mT)

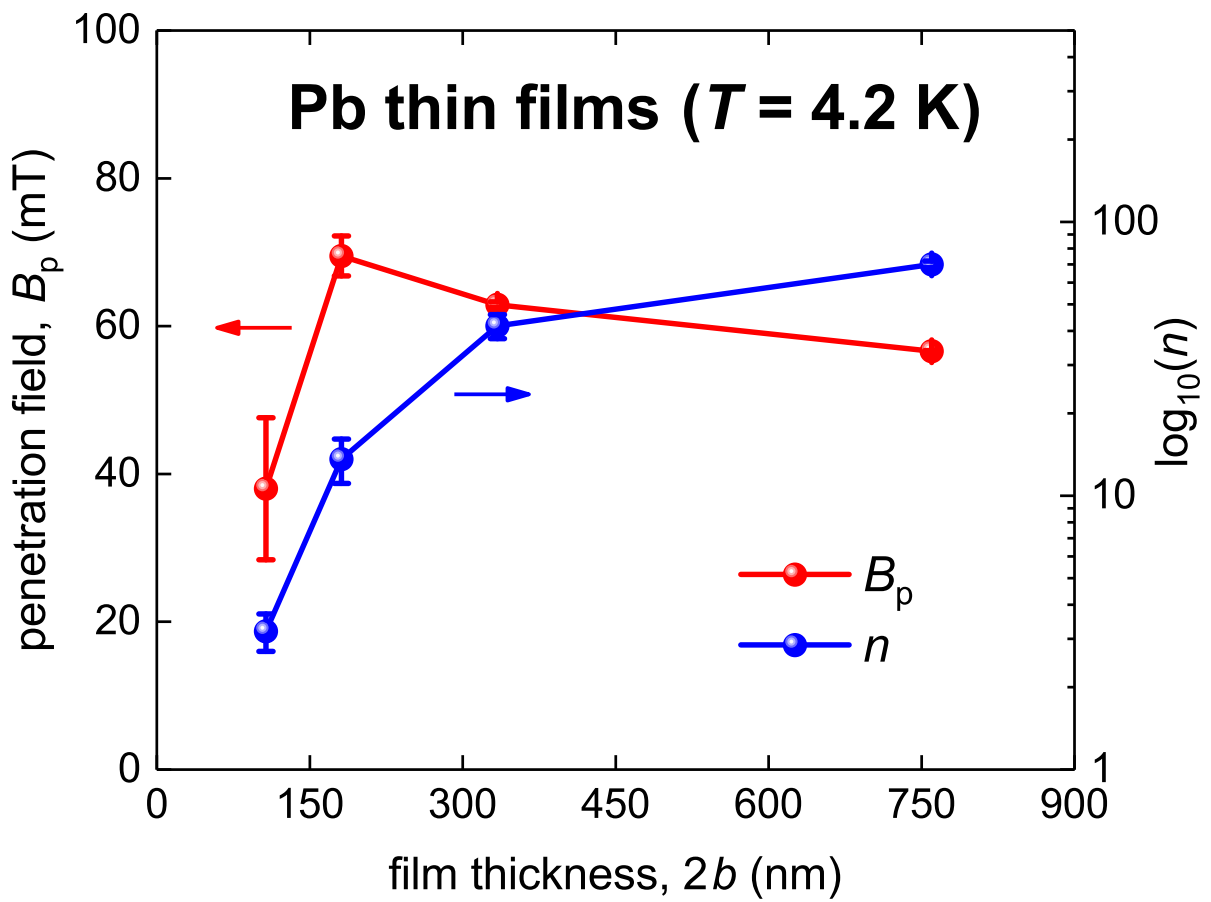

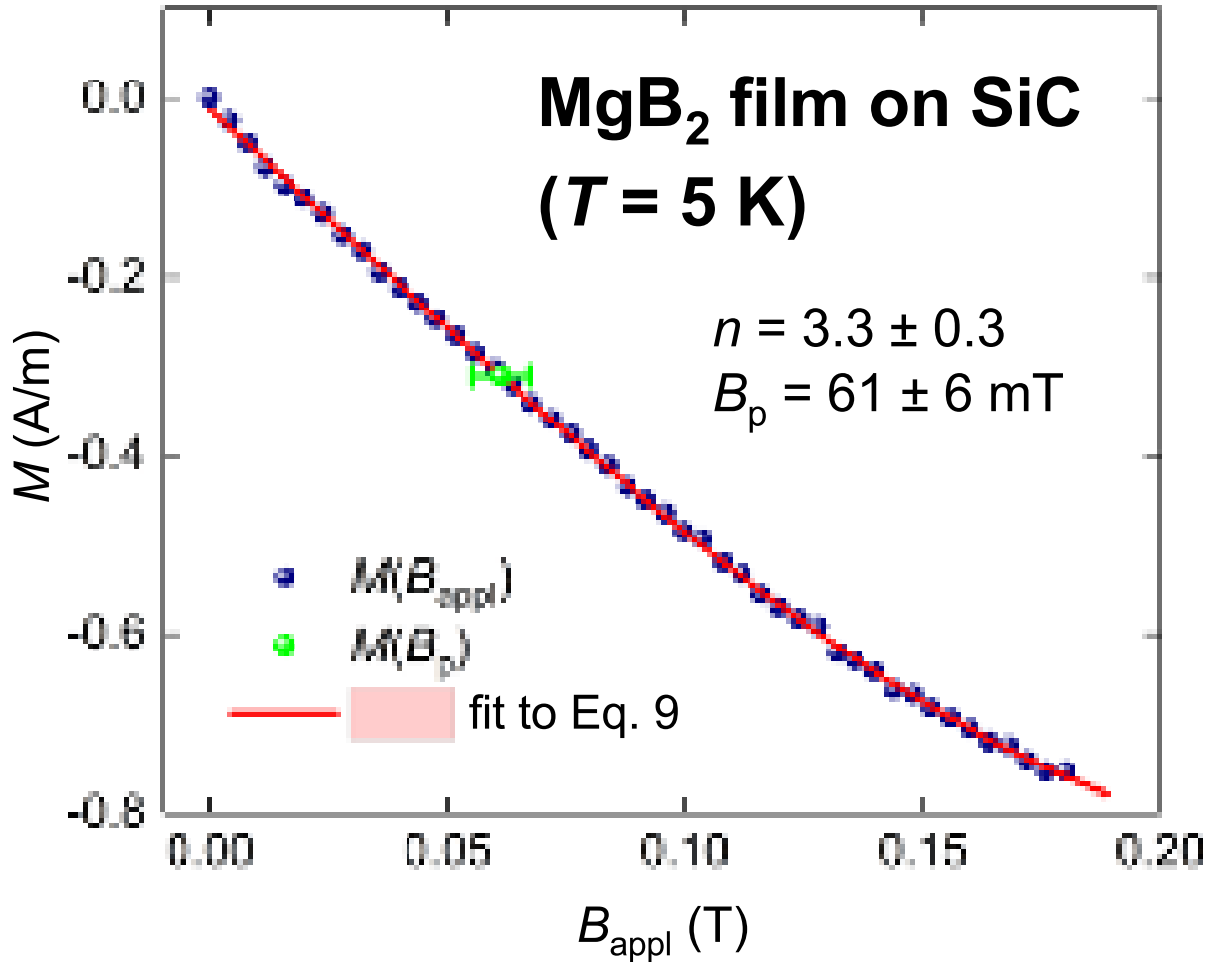

MgB2 film on SiC
(T = 5 K)
n = 3.3 ± 0.3
Bp = 61 ± 6 mT
M(Bappl)
M(Bp)
fit to Eq. 9
M (A/m)
0.0
-0.2
-0.4
-0.6
-0.8
0.00
0.05
0.10
0.15
0.20
Bappl (T)

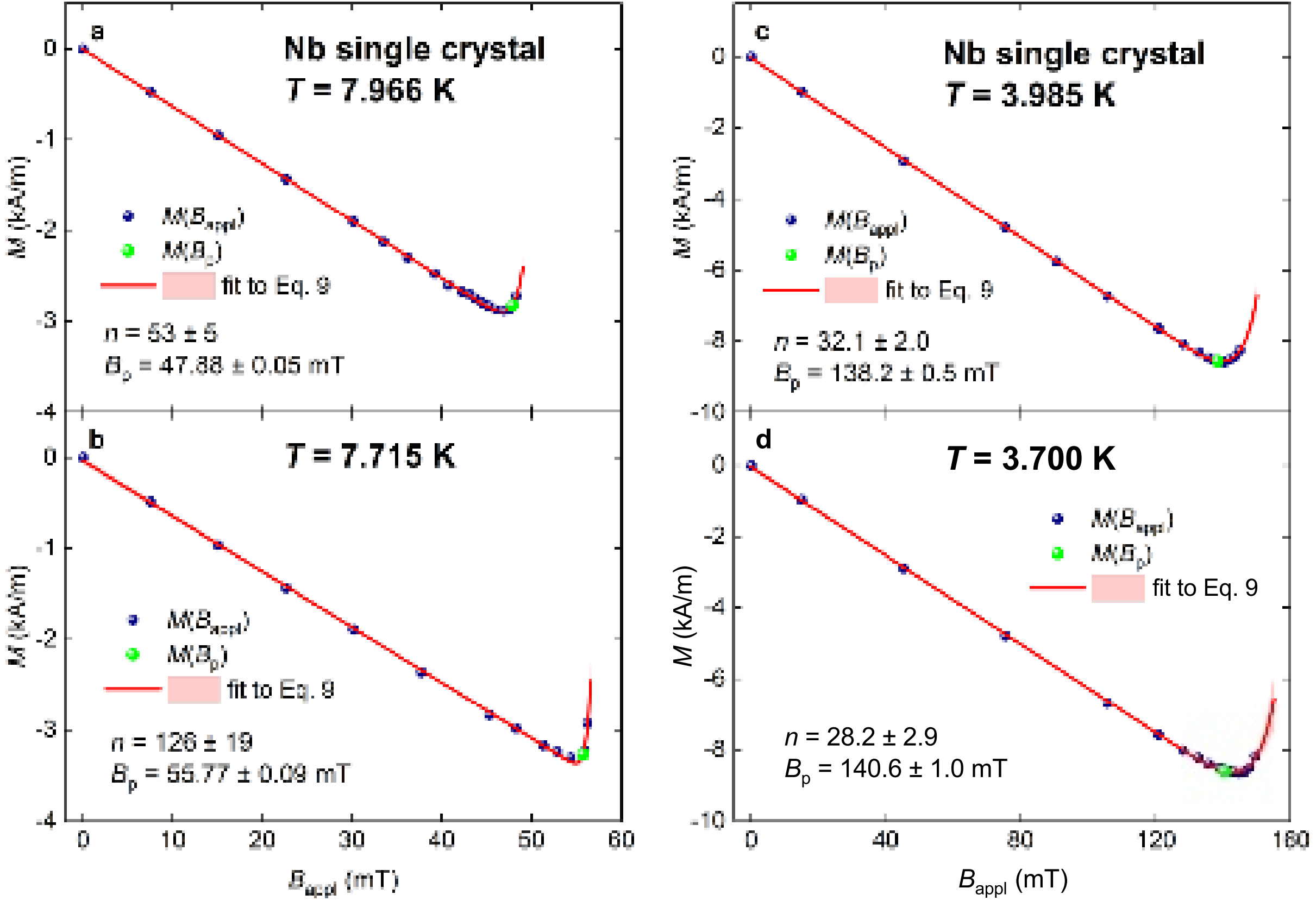

a
Nb single crystal
T = 7.966 K
M(B_appl)
M(B_p)
fit to Eq. 9
n = 53 ± 5
B_p = 47.88 ± 0.05 mT
b
T = 7.715 K
n = 126 ± 19
B_p = 55.77 ± 0.09 mT
c
Nb single crystal
T = 3.985 K
n = 32.1 ± 2.0
B_p = 138.2 ± 0.5 mT
d
T = 3.700 K
n = 28.2 ± 2.9
B_p = 140.6 ± 1.0 mT
M (kA/m)
B_appl (mT)

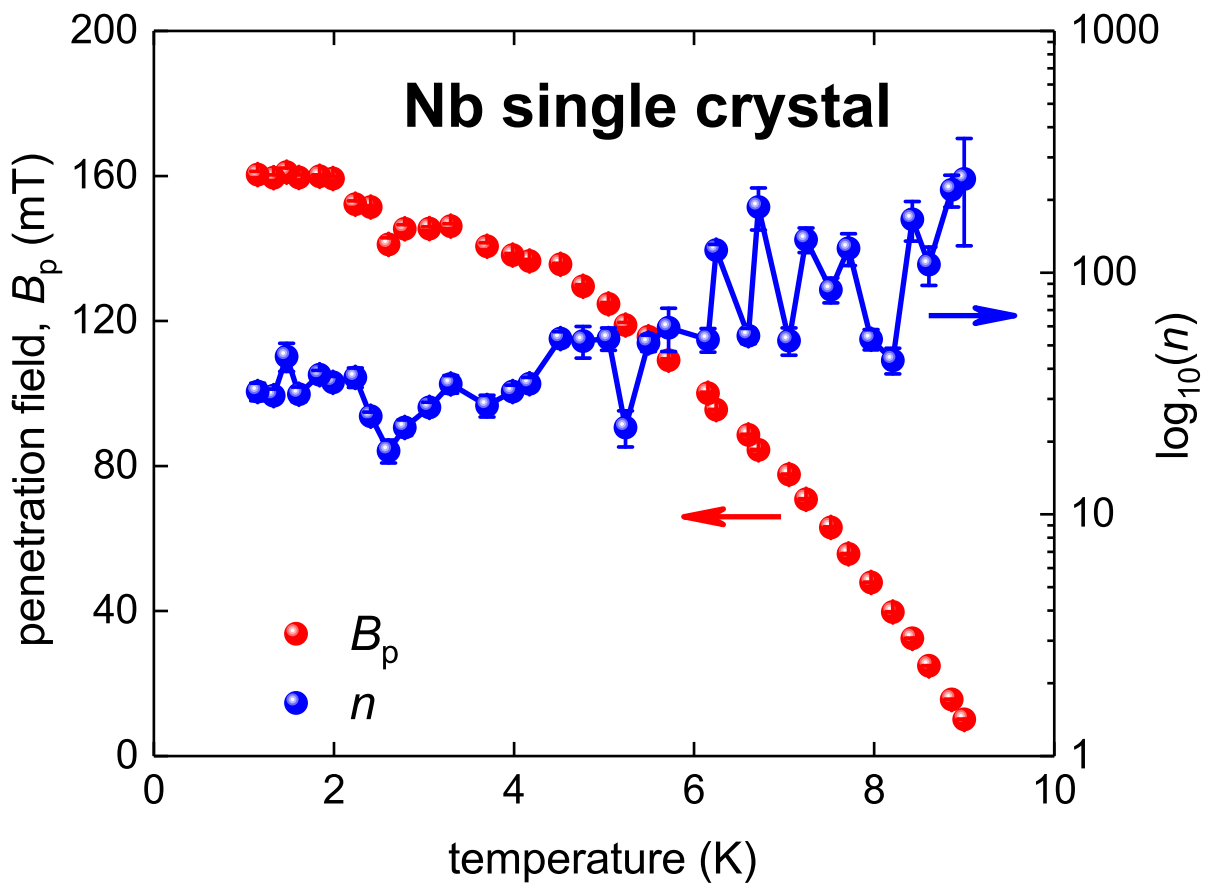

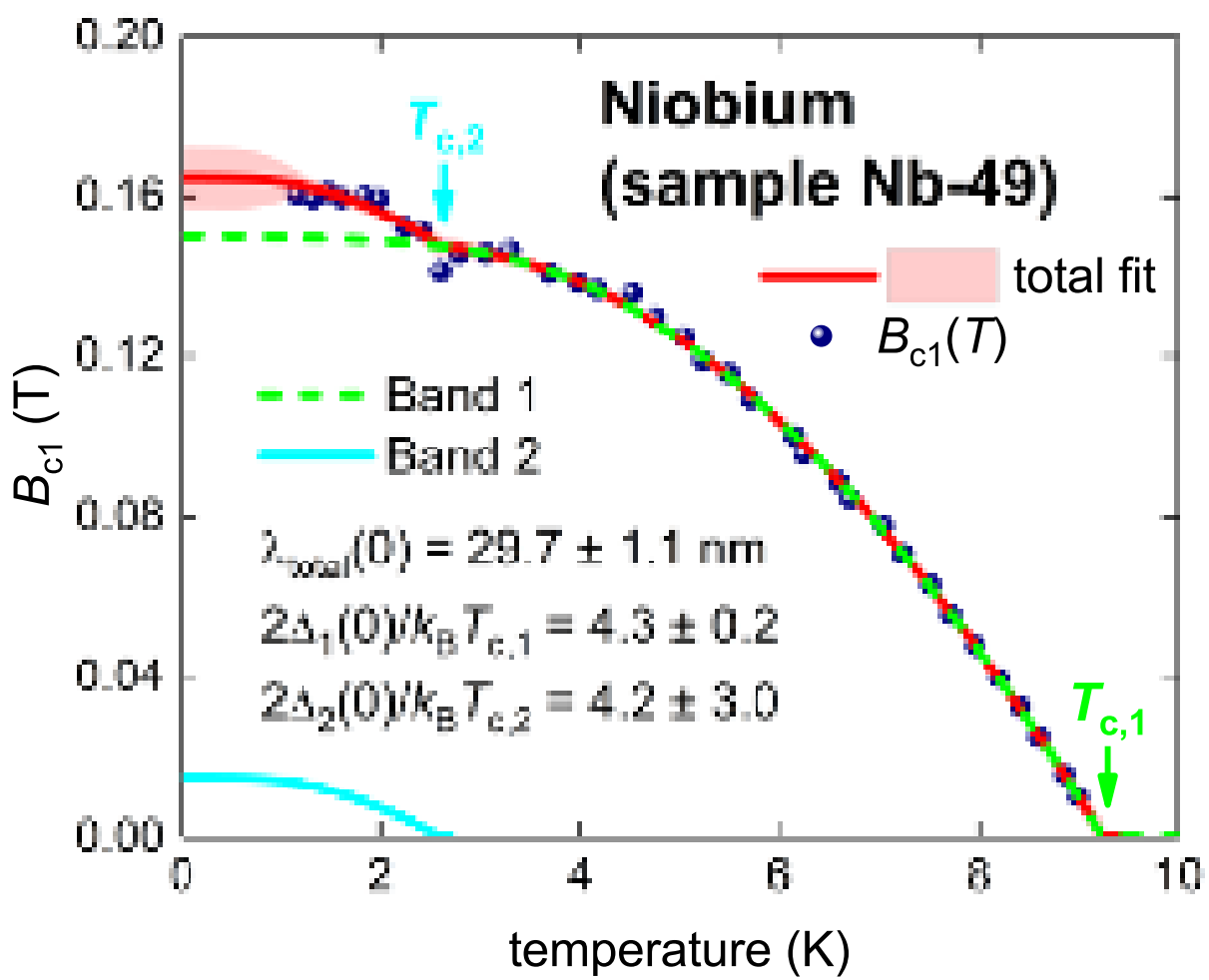

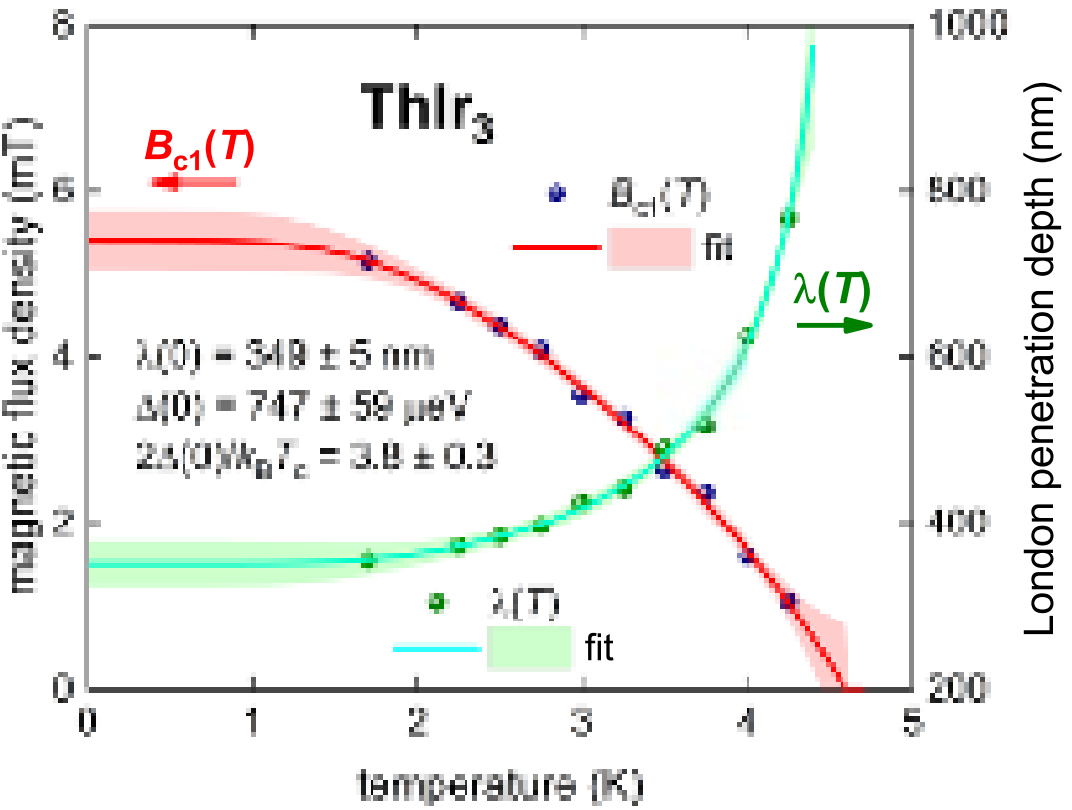
ThIr$_3$
$B_{c1}(T)$
$B_{c1}(T)$
fit
$\lambda(T)$
$\lambda(0)$ = 348 ± 5 nm
$\Delta(0)$ = 747 ± 59 μeV
$2\Delta(0)/k_B T_c$ = 3.8 ± 0.3
$\lambda(T)$
fit
magnetic flux density (mT)
London penetration depth (nm)
temperature (K)

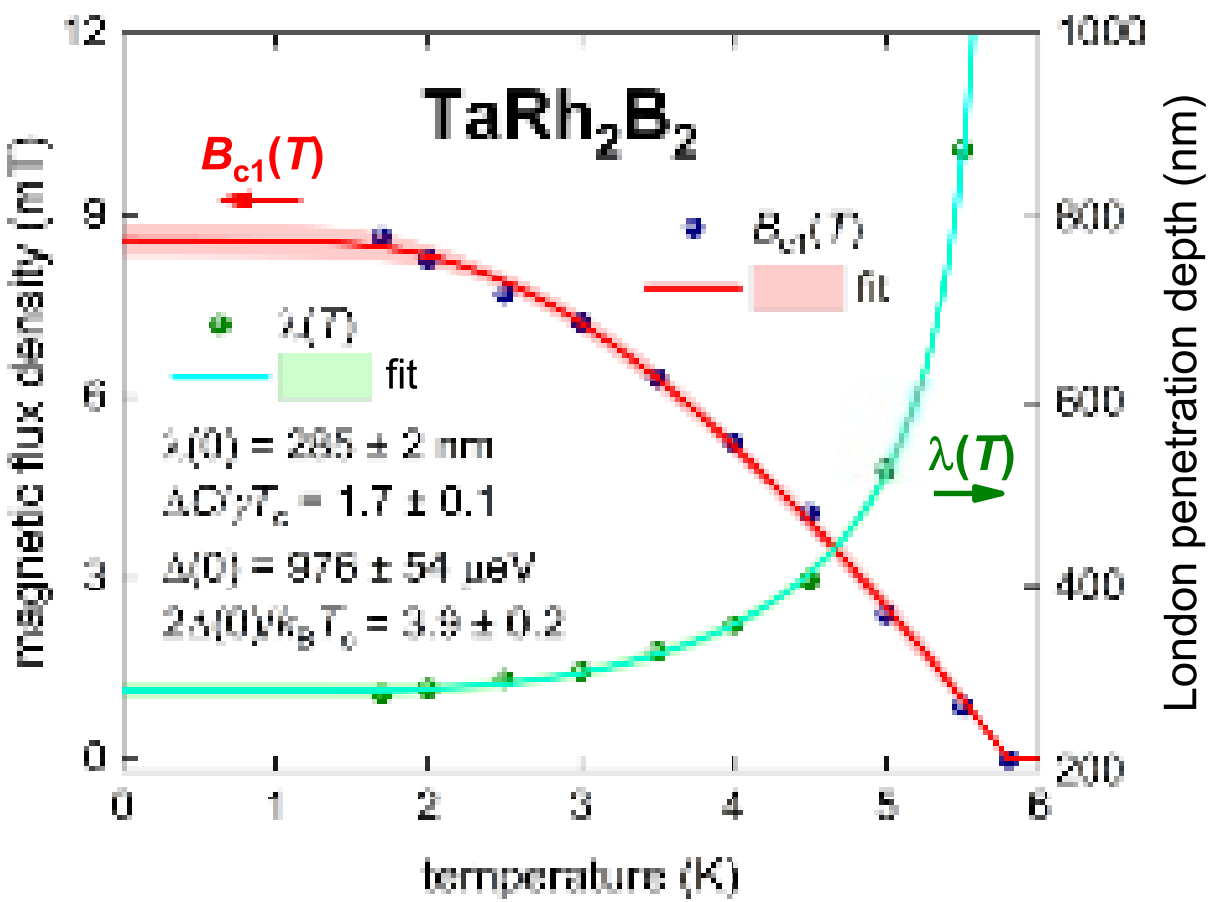

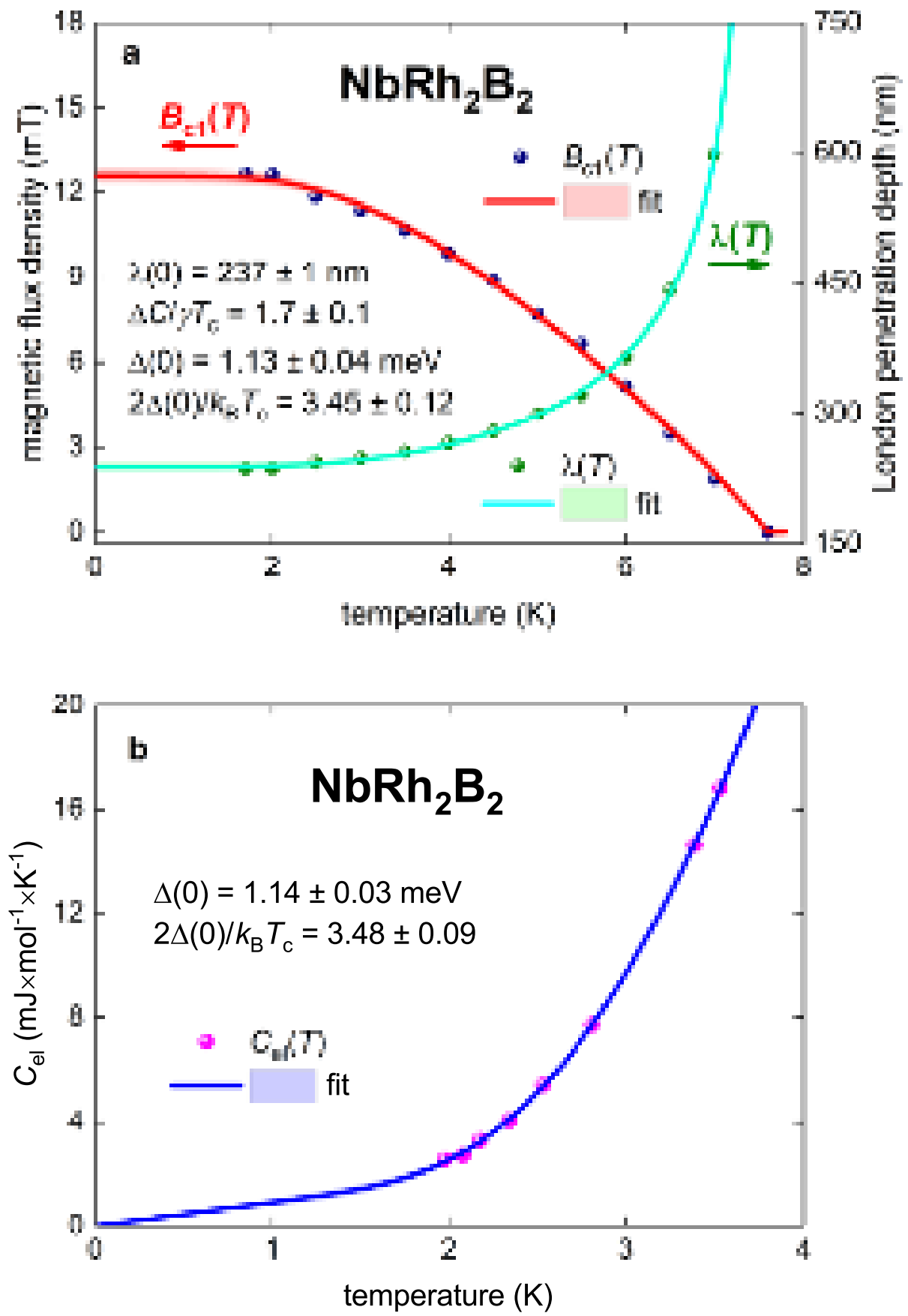
a
NbRh2B2
Bc1(T)
Bc1(T)
fit
λ(T)
λ(0) = 237 ± 1 nm
ΔC/γTc = 1.7 ± 0.1
Δ(0) = 1.13 ± 0.04 meV
2Δ(0)/kBTc = 3.45 ± 0.12
λ(T)
fit
magnetic flux density (mT)
London penetration depth (nm)
temperature (K)
b
NbRh2B2
Δ(0) = 1.14 ± 0.03 meV
2Δ(0)/kBTc = 3.48 ± 0.09
Cel(T)
fit
Cel (mJ×mol-1×K-1)
temperature (K)